\documentclass[aps,prc,nofootinbib,amsmath,amssymb]{revtex4}
\usepackage{eurosym}
\usepackage{amsmath,amssymb}
\usepackage{graphicx}
\usepackage{dcolumn}
\usepackage{bm}
\usepackage{color}
\usepackage{multirow}

\usepackage{float}
\usepackage{array}
\usepackage{dcolumn}
\usepackage{booktabs}
\usepackage[detect-all]{siunitx}
\usepackage{makecell}
\usepackage{xcolor}
\usepackage{yfonts}
\graphicspath{ {/Users/Nastia/Desktop/research_dis/papers_writing/xenes/nov_5} }

\newcolumntype{P}[1]{>{\centering\arraybackslash}p{#1}}
\begin{document}
\title{Effects of parallel electric and magnetic fields on Rydberg excitons
in buckled two-dimensional materials.}
\author{ Roman Ya. Kezerashvili$^{1,2}$ and Anastasia Spiridonova$^{1,2}$  }
\affiliation{ \mbox{$^{1}$Physics Department, New York City College
of Technology, The City University of New York,} \\
Brooklyn, NY 11201, USA \\
\mbox{$^{2}$The Graduate School and University Center, The
City University of New York,} \\
New York, NY 10016, USA}
\date{\today}

\begin{abstract}
We study direct and indirect magnetoexcitons in Rydberg states in monolayers and double-layer heterostructures of Xenes (silicene, germanene, and stanene) in external parallel electric and magnetic fields, applied perpendicular to the monolayer and heterostructure. We calculate binding energies of magnetoexcitons for the Rydberg states, 1$s$, 2$s$, 3$s$, and 4$s$, by numerical integration of the Schr\"{o}dinger equation using the Rytova-Keldysh potential for direct and both the Rytova-Keldysh and Coulomb potentials for indirect magnetoexcitons, respectively. Latter allows understanding a role of screening in Xenes. In the external perpendicular electric field, the buckled structure of the Xene monolayers leads to appearance of potential difference between sublattices allowing to tune electron and hole masses and, therefore, the binding energies and diamagnetic coefficients (DMCs) of magnetoexcitons. We report the energy contribution from electric and magnetic fields to the binding energies and DMCs. The tunability of the energy contribution of direct and indirect magnetoexcitons by electric and magnetic fields is demonstrated. It is also shown that DMCs of direct excitons can be tuned by the electric field, and the DMCs of indirect magnetoexcitons can be tuned by the electric field and manipulated by the number of hBN layers. Therefore, allowing the possibility of electronic devices design that can be controlled by external electric and magnetic fields and the number of hBN layers. The calculations of the binding energies and DMCs of magnetoexcitons in Xenes monolayers and heterostructures are novel and can be compared with the experimental results when they will be available.
\end{abstract}

\maketitle


\section{\label{sec:introduction} Introduction}
After graphene monolayer was successfully isolated and identified in 2004 \cite{Novoselov2004}, the rapid study of fundamental properties and applications in nano- and quantum devices of two dimensional materials has begun \citep{Mcdonell2016, Novoselov2016,Heerema2016,Iannaccone2018}. Initially, the main focus was graphene that led to the current vast body of knowledge about the structure, electric, magnetic, and transport properties of graphene \citep{CastroNeto2009a,Avouris2017}. Despite its uniqueness, graphene has a closed energy gap at $K$/$K^{\prime }$ points of the Brillouin zone complicating its application in electronics. As a result, the search and study of other 2D materials are underway.

Transition-metal dichalcogenides (TMDCs) and Xenes monolayers are among the main focus of the current research since TMDCs have the direct energy gap and Xenes have the energy gap that can be opened by the perpendicular electric field at $K$/$K^{\prime }$ points that can be utilized in the electronics.
In this paper, we adopt Xenes definition given in Ref. \cite{Molle2017, Zheng2020}. Xenes is a general name given to 2D buckled materials formed by elements from XIV group (First Generation \cite{Grazianetti2019}): silicene (Si), germanene (Ge), and stanene (Sn). In contrast to graphene, other elements from XIV group have the most stable monolayer form when atoms are arranged in the honeycomb pattern where lattices A and B are offset with respect to the plane of the monolayer \citep{Matthesparam,Molle2017}. The offset between latices is called the buckling constant or buckling factor, $d_0$. This unique structure of Xenes makes monolayers sensitive to the external electric field applied perpendicular to the monolayer allowing to control the band gap size.

The influence of the magnetic field on the excitonic wave function and binding energy of the exciton is well established. Excitons in a bulk, double quantum wells or 2D structures in the presence of the magnetic field result in a possible formation in these structures of quasiparticles called magnetoexcitons. Magnetoexcitons in TMDCs monolayers present a great interest since the binding energy of the magnetoexcitons can be tuned by the perpendicular to the monolayer magnetic field. On one hand, magnetoexcitons in TMDCs monolayers have been extensively studied in the past few years. For example, the exciton binding energies for Rydberg states are reported in Refs. \citep{Liu2019,Koperski, Stier_2016,Aivazian, Chen2019,Gor2019,Macneill, Striv,Ludwig,Plechinger,Zipfel, Kidd,Mayers, Goldstein, Robert, Liu_2020,Arora}, the Zeeman shift has been considered in Refs. \citep{DonckDM2018,Koperski, Stier_2016,Aivazian, Chen2019,Rybkovkiy, Gor2019,Macneill, Liu2019,Striv,Ludwig,Plechinger, Spiridonova, Stier2018}, while the diamagnetic shift was addressed in Refs. \citep{DonckDM2018,Liu2019, Stier_2016,Aivazian,Gor2019,Macneill,Plechinger,Stier2018,Luckert,donckexc, Choi2015,Walck,Han2018,Zipfel, Spiridonova}. On the other hand, there is a lack of research on magnetoexcitons in Xenes monolayers since the synthesis of Xenes monolayers has not been very successful compared to TMDCs because of Xenes instability in the air \cite{Tao2015}. In contrast to graphene, silicene monolayers do not occur in nature. However, silicene nanoribons were experimentally synthesized on a metal substrate \citep{Aufray2010, Padova2010} that opened the way for silicene, germanene, and stanene monolayers being transferred on metal \citep{Drummond2012,Saxena2016,Davila2014,Mannix2017,Grazianetti2019} and an insulating substrate such as MoS$_2$ \cite{Sante2019} and hexagonal boron nitride (hBN) \citep{LiWang2013, Khan2017}. Depositing Xenes on a metallic substrate is easier. For example, silicene can be grown on Ag (111) \citep{Tao2015,Ezawa1} or germanene synthesis by dry deposition on Au (111) surface \cite{Davila2014}. However, depositing Xene on a metal significantly alters properties of the Xene monolayer. Depositing Xenes on hBN is harder, but hBN preserves properties of the Xenes since Xenes and hBN weakly interact \cite{LiWang2013}. Properties of Xenes on different substrates are presented in Refs. \cite{Molle2017,LiWang2013}.

Nevertheless, magnetoexcitons in Xenes present a great interest for fundamental research and application in electronic devices. Out of all Xenes, silicene is most studied \cite{Zhao2016}. Due to Xenes monolayer unique buckling structure, electron-hole masses depend on the band gap that in return depends on the perpendicular to the monolayer electric field. Since the exciton binding energy depends on the reduced mass of the exciton, by applying the electric field the exciton binding energy can be tuned.

Optical properties of Xenes have been addressed in Refs. \citep{Matthesparam,Bechstedt2012,Fadaie2016,Stille2012}, and magneto-optical properties are studied in Ref. \cite{Muoi2020,Chowdhury2016}. Hall effect \cite{Ezawa1}, the valley-locked spin-dependent Seeback effect \cite{Zhai2018}, anomalous quantum Hall effect \cite{Ezawa2}, and quantum spin Hall effect \cite{Zhao2020} are being addressed since they are essential for the use of Xenes in nano- and quantum  devices \citep{Glavin2020, Lyu2019, Tao2020, Chen2018, Zhao2020}. Currently, there is a lack of studies on magnetoexciton binding energies, Zeeman shift, and diamagnetic coefficients (DMCs) in Xenes monolayers. However, the Landau levels have been addressed in Refs. \citep{Ezawa1,Muoi2020, Tsaran2014}. Moreover, a particular interest presents the study of magnetoexcitons formation in a double-layer structure of Xenes separated by layers of hBN. In such system electrons are confined in one Xene  monolayer, while an equal number of positive holes are located in a parallel Xene monolayer at a distance $D$ away. Thus, the electron and hole are spatially separated by a dielectric. The system of the charge carriers in two parallel Xenes layers can be treated as a two-dimensional system without interlayer hopping. In this system, the electron-hole recombination due to the tunneling of electrons and holes between different Xenes monolayers is suppressed by the dielectric barrier produced by hBN monolayers that separates  the Xenes monolayers. Therefore, the indirect (dipolar) excitons, formed by electrons and holes, located in two different Xenes monolayers, have a longer lifetime than the direct magnetoexcitons \cite{Lozovik1976}. In the past 6 years binding energies of Rydberg states of excitons in TMDCs monolayers and heterostructures in the magnetic field have been intensively studied experimentally and theoretically \citep{Liu2019,Koperski, Stier_2016,Aivazian, Chen2019,Gor2019,Macneill, Striv,Ludwig,Plechinger,Zipfel, Kidd,Mayers, Goldstein, Robert, Liu_2020,Arora}, and the diamagnetic shifts were considered in Refs. \citep{DonckDM2018,Liu2019, Stier_2016,Aivazian,Gor2019,Macneill,Plechinger,Stier2018,Luckert,donckexc, Choi2015,Walck,Han2018,Zipfel, Spiridonova}. This motivates us to focus our study solely on the $s$ Rydberg states of excitons in Xenes in the magnetic field.

In this paper, we study the dependence of the magnetoexciton binding energy of Rydberg states, 1$s$, 2$s$, 3$s$, and 4$s$, on the perpendicular to the monolayer electric and magnetic fields and calculate the diamagnetic coefficients for (i) the direct $A$ magnetoexcitons in Xenes monolayers and (ii) for the indirect $A$ magnetoexcitons in Xenes heterostructure formed by two monolayers of the same Xene and separated by $N$ monolayers of hBN. The heterostructure is denoted as X-hBN-X, and the number of hBN layers is varied between 1 and 6. We numerically solve the Schr\"{o}dinger equation for the magnetoexciton in external electric and magnetic fields to obtain eigenfunctions and eigenvalues. Then we calculate the energy contribution from the magnetic field to the binding energy, as well as its dependence on the applied electric field. From the energy contribution from the magnetic field to the binding energy, we extract the DMCs. For the direct exciton, we solve the Schr\"{o}dinger equation with the Rytova-Keldysh potential \citep{Rytova, Keldysh}, and for the indirect exciton, we solve the Schr\"{o}dinger equation with the Rytova-Kyldysh and the Coulomb potentials. This allows to understand the role of screening in Xenes. So, we study the dependence of the binding energy of indirect magnetoexcitons on the external electric and magnetic fields, as well as on the separation of the Xenes layers by the layers of hBN. The calculations of the binding energies of magnetoexcitons and DMCs in Xenes monolayers and heterostructure are reported for the first time. As a result, our study demonstrate a tunability of the energy contribution of direct and indirect magnetoexcitons by electric and magnetic fields. It is shown that DMCs of direct
excitons can be tuned by the electric field, while the DMCs of indirect magnetoexcitons can be tuned
by the electric field and manipulated by the number of hBN layers. Therefore, we demonstrate a
possibility of electronic devices design that can be controlled by the external electric and magnetic fields
and the number of hBN layers.

The paper is organized in the following way. In Sec.~\ref{sec:theory} is given the theoretical formalism for the description of an electron and hole in buckled 2D materials. Here we present the
effective mass approach for magnetoexcitons in buckled 2D materials and the Schr\"{o}dinger equation for the exciton in parallel electric and magnetic fields, which are perpendicular to the Xene monolayer.
The theoretical consideration of indirect magnetoexcitons in Xenes heterostructure is given in Subsec.~\ref{ssec:indirect}. The results of calculations of the energy contribution from the electric and magnetic fields  to the binding energy and diamagnetic shifts of the direct and indirect magnetoexcitons and dependence of the energy contribution and diamagnetic shifts for indirect magnetoexcitons in Xenes heterostructures on the number of hBN layers on the external electric field are presented in Sec. ~\ref{sec:results}. Conclusions follow in Sec.~\ref{sec:conclusion}.

\section{\label{sec:theory} Theoretical Formalism}

\subsection{\label{ssec:theoryhamiltonian}Electron and hole in buckled 2D
materials}

Let us provide an outline of the low-energy model that describes excitons
states in Xenes monolayers and heterostructures under applied electric and
magnetic fields. We are considering an electron and hole in parallel magnetic $\mathbf{B}%
(0,0,B_{z})\equiv $ $\mathbf{B}(0,0,B)$ and electric $\mathbf{E}%
(0,0,E_{z})\equiv \mathbf{E}(0,0,E)$ fields, which are perpendicular to a Xene monolayer or heterostructure as it is shown in Fig. 1. Monolayers of silicene, germanene, and low-buckled stanene
can be pictured as honeycomb graphene monolayers but have an out of plane
buckling such that the A and B triangular sublattices are offset with
respect to the plane of the monolayer and separated
by a distance $d_{0}$. This distance is known as the buckling constant or buckling factor.
The intrinsic sensitivity of Xenes to an external electric field applied
perpendicular to the plane of the monolayer is mounted due to the offset between the two triangular sublattices. In particular, this asymmetry
causes an on-site potential difference to occur between sublattices when an
out-of-plane electric field is applied. The band structure of Xenes in the vicinity of the $K/K^{\prime }$ points
resembles graphene when there is no external electric field; though, the intrinsic gaps of Xenes are significantly
larger than that of graphene. The application of a perpendicular electric
field creates a potential difference between the sublattices, causing a
change in the band gap in the monolayer Xenes, which in turn changes the effective
masses of the electrons and holes.
\begin{figure}[t]

\begin{centering}
\includegraphics[width=16.0cm]{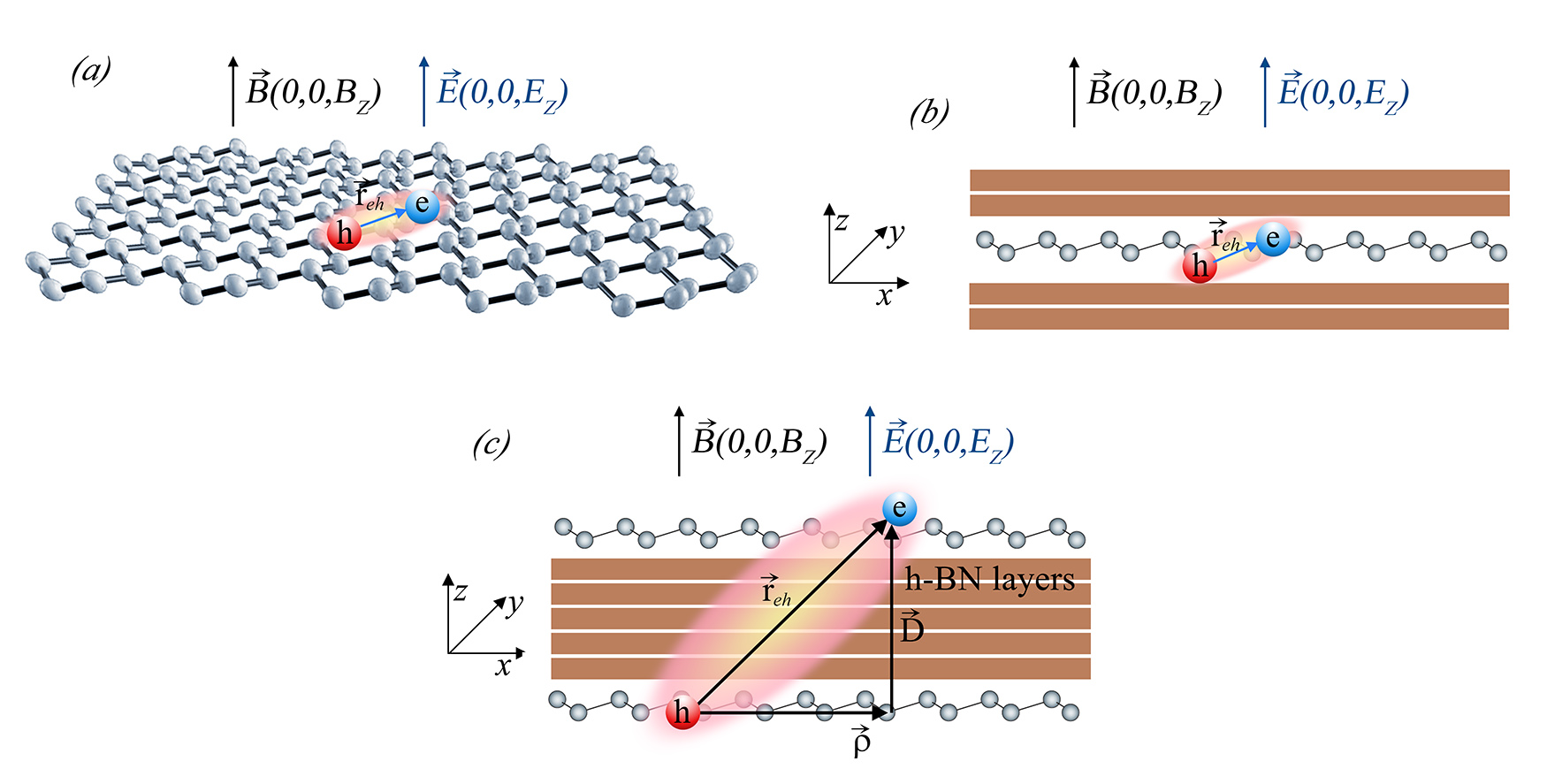}
\caption{(Color online) Schematics for magnetoexcitons in Xenes monolayer and heterostructure. $(a)$ A direct magnetoexciton in the freestanding buckled honeycomb lattice structure of silicene monolayer. $(b)$ A magnetoexciton in silicene encapsulated monolayer. $(c)$ An indirect magnetoexciton in silicene heterostructure.}
\label{fig1M}
\end{centering}
\end{figure}

The single-particle spectrum of electronic states in monolayer Xenes in the
electric field acting along the $z$-axis in the vicinity of the $K/K^{\prime
}$ points is described by a two-dimensional (2D) massive Dirac Hamiltonian is
given in Ref. \cite{Tabert} as ($\hbar =c=1)$:
\begin{equation}
\hat{H}_{0}=v_{F}\left( \xi p_{x}\hat{\tau}_{x}+p_{y}\hat{\tau}_{y}\right)
-\xi \Delta _{so}\hat{\sigma}_{z}\hat{\tau}_{z}+\Delta _{z}\hat{\tau}_{z}.
\label{eq:taberthamiltonian}
\end{equation}%
In Hamiltonian \eqref{eq:taberthamiltonian} $v_{F}$ is the Fermi velocity, $%
p_{x}$ and $p_{y}$ are the components of momentum in the monolayer $xy$-plane,
relative to the $K$ points, $\hat{\tau}$ and $\hat{\sigma}$ are
the pseudospin and real spin Pauli matrices, respectively, $2\Delta _{so}$
is the intrinsic band gap, $\xi ,\sigma =\pm 1$ are the valley and spin
indices, respectively, and $\Delta _{z}=ed_{0}E_{z}\equiv $ $ed_{0}E$ is the gap induced
by the external electric field $E_{z}\equiv E$, acting along the $z$-axis, where $d_{0}$
in the latter expression is the buckling constant. As it follows from Refs. \cite{Ezawa2,Tabert,Ezawa3}, Hamiltonian \eqref{eq:taberthamiltonian}, which describe electronic states in Xenes, is analogous to the 2D Dirac Hamiltonian for TMDCs monolayers. The first term in Eq.~%
\eqref{eq:taberthamiltonian} is the same as that of the low-energy
Hamiltonian in graphene~\cite{CastroNeto2009a,Abergel2010}. The last term in %
\eqref{eq:taberthamiltonian} describes the sublattice potential difference
that could arise from the application of a perpendicular electric field \cite%
{Drummond2012, Ezawa1, Ezawa2, Ezawa3}, while the spin-orbit coupling~\cite%
{Kane2005} with an intrinsic band gap of $2\Delta _{so}$ is given by the
second term. The Hamiltonian for the single-particle spectrum of electronic
states in monolayer Xenes can be obtained by replacing $\mathbf{p}%
\rightarrow \mathbf{p}+e\mathbf{A}$, where $\mathbf{A}$ is the vector
potential such that $\mathbf{A}=\nabla \times \mathbf{B}$, where $\mathbf{B}$
is the magnetic field.
\begin{figure}[b]
\begin{centering}
\includegraphics[width=12.0cm]{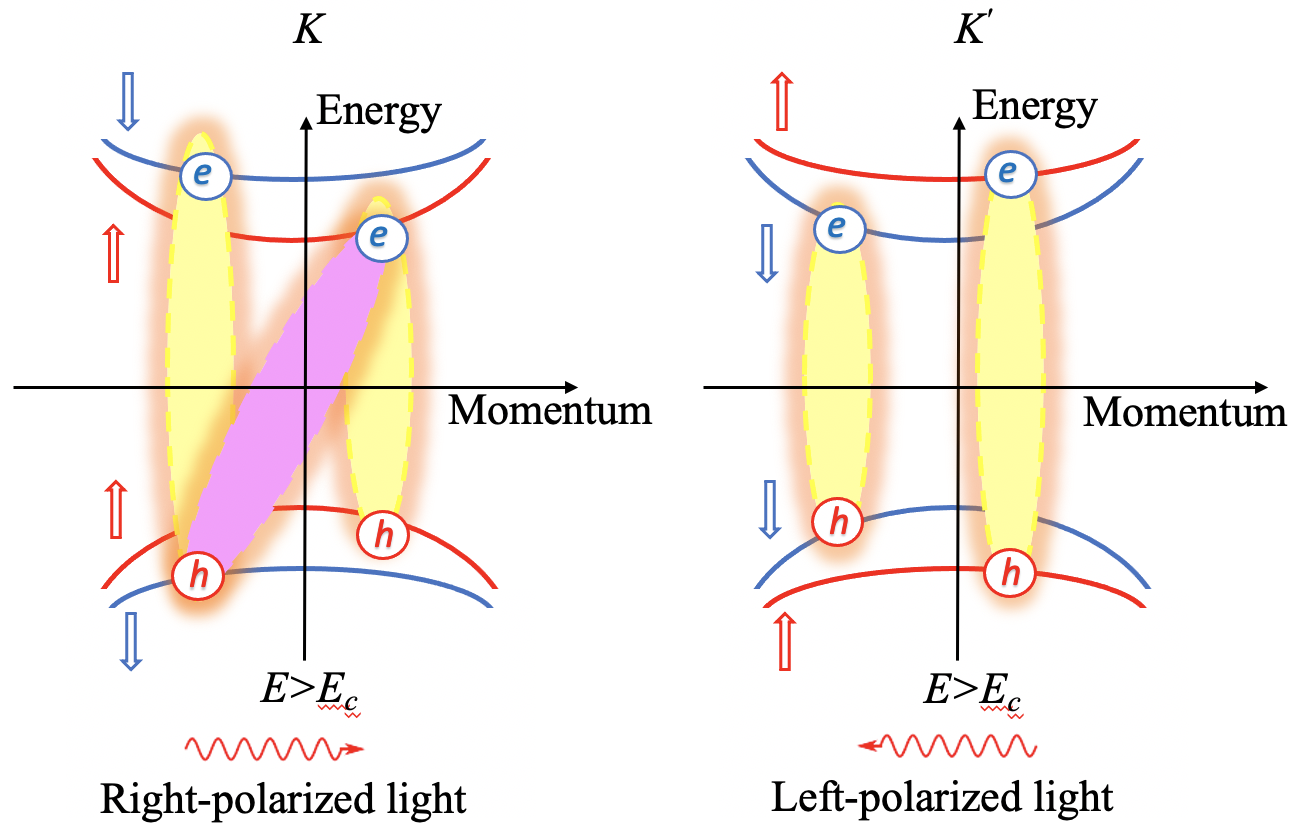}
\caption{Schematic band structure and electronic dispersions in the silicene monolayer for bright and dark excitons in the $K$ and $K'$ valleys when the electric field, $E$, is perpendicular to the monolayer. The pseudospin is opposite to spin at the $K$ point. Spin-up and spin-down bands are denoted by red and blue lines, respectively. The yellow shadowed ovals are the bright excitons and correspond to the lowest optically induced transition between the bands of the same spin at the $K$ and $K'$ point. The magenta shadowed oval is the spin-forbidden dark exciton (the second one is not shown). The units of the vertical and horizontal axes are arbitrary. At point $K$ the right circular polarized light couples to both $A$ and $B$ exciton transitions. At point $K'$  the left circular polarized light couples to $A$ and $B$ excitons.}
\label{Xenesexciton}
\end{centering}
\end{figure}
Using Eq.~\eqref{eq:taberthamiltonian} one may write the low-energy
eigenvalues for charge carriers near the $K/K^{\prime }$ points as \cite%
{Tabert}:
\begin{equation}
E(k)=\sqrt{\Delta _{\xi \sigma }^{2}+v_{F}^{2}p^{2}},  \label{eq:disprel}
\end{equation}%
where
\begin{equation}
\Delta _{\xi \sigma }=|\xi \sigma \Delta _{so}-ed_{0}E|
\label{eq:deltaez}
\end{equation}%
is the electric field-dependent band gap at $p=0$. We note that when $E=0$,
the spin-up and spin-down bands of the valence and conduction
bands are degenerate. In other words, spin-orbit splitting only manifests
itself at non-zero external electric fields. At non-zero electric fields,
both the valence and conduction bands split, into upper bands with a large
gap (when $\xi =-\sigma $), and lower bands with a small gap (when $\xi
=\sigma $). We call the excitons formed by charge carriers from the large
gap $A$ excitons, and those formed by charge carriers in the small gap $B$
excitons. The direct bright $A$ and $B$ excitons consist of the electron and hole with parallel spins \citep{Echeverry2016, Malic2018}. There exist two additional states when the electron and hole have antiparallel spins, but these excitons are optically forbidden and called spin forbidden dark excitons. The schematic band structure of $A$ and $B$ excitons formation in $K$ and $K'$ valleys under the right- and left-polarized light, respectively, is given in Fig. \ref{Xenesexciton}. Following Ezawa \cite{Ezawa3}, we show the case when the applied perpendicular electric field is bigger than the critical value of the electric field, $E_c$. When the external field reaches a critical value $E_{c}=\Delta
_{so}/(ed_{0})$, the lower bands form a Dirac cone at the $K/K^{\prime }$
points. The corresponding values of the critical electric field for
monolayer Xenes are given in Ref. \cite{BBKPRB2019}.
The conduction and valence bands are parabolic in the vicinity of the $%
K/K^{\prime }$ points. One
can find the effective mass of charge carriers near the $K/K^{\prime }$
points as $m=\Delta _{\xi \sigma }/v_{F}^{2}$ \citep{Pan2015,Matthesparam}. The effective masses
of electrons and holes are the same due to the symmetry between the lowest
conduction and highest valence bands, and can be written as a function of
the external electric field in the following form:
\begin{equation}
m=\frac{\lvert \xi \sigma \Delta _{so}-ed_{0}E\rvert }{%
v_{F}^{2}}.  \label{eq:effmassEz}
\end{equation}%

Therefore, it is worth mentioning that the reduced mass of an exciton in
Xenes $\mu =$ $m/2$ is a function of the external electric field $E$ for $A$ and $B$
excitons. Following \textit{ab initio} calculations~\cite{Drummond2012},
which determined that the crystal structure of silicene becomes unstable
around 2.6 V/{\AA}, we consider in our calculations external electric fields up to 2.7
V/\AA . Analysis of Eq. (\ref{eq:effmassEz}) shows that the value of reduced
mass depends on the band gap, Fermi velocity, and buckling constant. For example,
for freestanding silicene, germanene, and stanene 2$\Delta _{so}$ is 1.9 (1.55) meV \cite{Matthesparam} (\cite{LiuFeng2011}),
33 (23.9) meV \cite{Matthesparam} (\cite{LiuFeng2011}), and 101 (73.5) meV \cite{Matthesparam} (\cite{Liu2011}), while the
corresponding Fermi velocities are 6.5 (5.3)$\times $10$^{5}$ m/s \cite{Matthesparam} (\cite{Matthes2013}), 6.2 (5.2)$\times $10$^{5}$ m/s
\cite{Matthesparam} (\cite{Matthes2013}), and 5.5 (4.8 )$\times $10$^{5}$ m/s \cite{Matthesparam} (\cite{Liu2011}).
Let us notice that the citations in parenthesis are related to the data in parentheses. At small
electric fields, germanene and especially stanene show significant
differences between the reduced masses of the $A$ and $B$ excitons. The latter
is due to their large intrinsic band gaps. Silicene, which has an intrinsic
band gap on the order of a couple of meV, exhibits very little difference
between the reduced masses of $A$ and $B$ excitons, even at relatively small
electric fields. At large electric field the difference between the $A$ and
$B$ exciton reduced mass is negligible in silicene and germanene. In all
cases, the mass of the $A$ exciton exceeds the mass of the $B$ exciton.

\subsection{\label{ssec:theoryexcitons}Effective mass approach for
magnetoexcitons in buckled 2D materials}

Excitonic systems are many-body systems, and the most systematic approach requires the
use of quantum field theory. However, these excitonic systems can be well
approximated and treated in the framework of few-body physics. There are different
approaches to solving the two-body eigenvalue and eigenfunction problem in two-dimensions for
interacting electrons and holes \cite{RK2019}. One can
start from the effective low-energy single-electron Hamiltonian. As it is mentioned above, we consider an electron and hole in parallel magnetic $\mathbf{B}(0,0,B)$ and electric $\mathbf{E}(0,0,E)$ fields, which are perpendicular to a monolayer. Using the
low-energy effective two-band single-electron Hamiltonian in the form of a
spinor with a gapped spectrum in the $\bm{k}\cdot \bm{p}$ approximation one obtains
the two-particle Dirac type equation. In the case of Xenes starting from
Hamiltonian\ \eqref{eq:taberthamiltonian} for two independent particles with
masses $m_{j}$ and coordinates $\mathbf{r}_{j}$ we obtain product states
with single particle wave function $\varphi _{1,j}(\mathbf{r}_{j})$ of energy
$\mathfrak{e}_j$ from Dirac type equation that reads:
\begin{equation}
v_{F}\left(
\begin{array}{cc}
-\sigma \xi \Delta _{so}/v_{F}+ed_{0}E/v_{F} & \xi \partial
_{x_{j}}+A_{x}(\mathbf{r}_{j})-i\partial _{y_{j}}-iA_{y}(\mathbf{r}_{j}) \\
\xi \partial _{x_{j}}+A_{x}(\mathbf{r}_{j})+i\partial _{y_{j}}+iA_{y}(%
\mathbf{r}_{j}) & \sigma \xi \Delta /v_{F}-ed_{0}E/v_{F}%
\end{array}%
\right) \left(
\begin{array}{c}
\varphi _{1,j}(\mathbf{r}_{j}) \\
\varphi _{2,j}(\mathbf{r}_{j})%
\end{array}%
\right) =\mathfrak{e}_j\left(
\begin{array}{c}
\varphi _{1,j}(\mathbf{r}_{j}) \\
\varphi _{2,j}(\mathbf{r}_{j})%
\end{array}%
\right).
\label{eq:Dirac}
\end{equation}
In Eq. (\ref{eq:Dirac}) $A_x$ and $A_y$ are the $x$ and $y$ components of the vector potential. The term related to the electric field vanishes because
it is directed along $z$-axis. However, the effect of the electric field action is present through the effective mass term $ed_{0}E/v_{F}$.
The component $\varphi _{1,j}$ of the Dirac spinor satisfies the following equation
\begin{equation}
\frac{1}{2m_{j}}[(i\partial _{x_{j}}+A_{x,j})^{2}+(i\partial
_{y_{j}}+A_{y,j})^{2}]\varphi _{1,j}=\frac{\mathfrak{e}%
_{j}^{2}+v_{F}^{2}B-(\sigma \xi \Delta _{so}+ed_{0}E)^{2}}{%
2v_{F}^{2}m_{j}}\varphi _{1,j}  \label{schroed0}
\end{equation}%
and the second component of the spinor is related to the first as
\begin{equation}
\varphi _{2,j}(\mathbf{r}_{j})=\frac{v_{F}}{\mathfrak{e}_{j}+v_{F}(\sigma \xi
\Delta _{so}+ed_{0}E)^{2}}\left[ i\partial _{x_{j}}+A_{x}(\mathbf{r%
}_{j})-\partial _{y_{j}}+iA_{y}(\mathbf{r}_{j})\right] \varphi _{1,j}(%
\mathbf{r}_{j})\ .
\end{equation}%
This implies that the eigenvalue of the Dirac equation reads $\mathfrak{e}%
_{j}=\sqrt{2m_{j}v_{F}^{2}E_{j}- v_{F}^{2}B+(2\Delta
_{so}+ed_{0}E)^{2}}$, where $E_{j}$ is the eigenvalue of the
corresponding equation (\ref{schroed0}).  Under the
assumption that the electron and hole bands are isotropic and parabolic,
which is a good approximation for the low-energy spectrum of 2D materials,
this Hamiltonian implies that both the electron and hole single
particle states form a single parabolic band. The corresponding eigenproblem
equation reduces to the Schr\"{o}dinger equation in the effective mass
approximation. This approach is common in the literature to describe
excitons in 2D materials. See, for example, Refs.~~\cite%
{Berkelbach2013,Fogler2014,Berman2016,RKPRA2013,Urbaszek2018}. We cite these works, but the recent literature on the subject is not limited
by them. Below, we
follow the effective mass approximation and, therefore, considering the Mott-Wannier excitons \cite{Wannier}.

To find the eigenfunctions and eigenenergies of a 2D exciton in
Xenes in external parallel electric and magnetic fields, we write the Schr\"{o}dinger
equation for an interacting electron and hole. Because we are considering
the varying electric field $E$, which is directed along $z$-axis, the
corresponding term in the 2D Schr\"{o}dinger equation vanishes. However,
the effect of the electric field action is present through the effective
mass. Thus one can write 2D Schr\"{o}dinger equation for the interacting
electron-hole system in the following form \cite{Herold}:
\begin{equation}
\left[ \frac{1}{2m_{e}}\left( -i\nabla _{e}+e\mathbf{A(r}_{e})\right) ^{2}+%
\frac{1}{2m_{h}}\left( -i\nabla _{h}-e\mathbf{A(r}_{h})\right) ^{2}+V\left(
r_{e},r_{h}\right) \right] \psi \left( \mathbf{r}_{e},\mathbf{r}_{h}\right)
=\mathcal{E}\psi \left( \mathbf{r}_{e},\mathbf{r}_{h}\right) ,  \label{eq:ehschro}
\end{equation}%
where $e$ and $h$ are the indices referring to the electron and hole,
respectively,  $\mathbf{r}_{e}$ and $\mathbf{r}_{h}$ are 2D coordinates of
the electron and hole, $m_{e}$ and $m_{h}$ are the masses of charge
carriers given by Eq.~\eqref{eq:effmassEz}, $\mathbf{A(r}_{e(h)})=\mathbf{%
B\times r}_{e(h)}/2$ is a gauge vector potential, $V\left(
r_{e},r_{h}\right) $, is the potential of interaction between the electron and
hole confined in 2D space.
The latter interaction between the electron and hole
in three-dimensional (3D) homogeneous dielectric environments is
described by the Coulomb potential, but this interaction is modified in 2D monolayer. This modification is
a result of the reduced dimensionality and includes screening effects. The corresponding potential was first derived in Ref.~\cite%
{Rytova} and a decade later was independently obtained in Ref.~\cite{Keldysh}%
--{} we refer to it as the Rytova-Keldysh (RK) potential. For almost a decade the celebrated Rytova-Keldysh potential has been widely used to describe
electromagnetic interaction between charged carriers of few-body complexes in transition-metal dichalcogenides, phosphorene, and Xenes monolayers. The RK potential is a central
potential and the interaction between the electron and hole for direct
excitons in a Xene monolayer has the form \cite{Rytova, Keldysh}:
\begin{equation}
V_{RK}(r)=-\frac{\pi ke^{2}}{2\kappa \rho _{0}}\left[ H_{0}\left( \frac{r}{\rho
_{0}}\right) -Y_{0}\left( \frac{r}{\rho _{0}}\right) \right] ,  \label{eq:rk}
\end{equation}%
where $r=r_{e}-r_{h}$ is the relative coordinate between the electron and
hole. In Eq.~(\ref{eq:rk}), $e$ is the charge of the electron, $\kappa
=(\epsilon _{1}+\epsilon _{2})/2$ describes the surrounding dielectric
environment, $\epsilon _{1}$ and $\epsilon _{2}$ are the dielectric
constants below and above the monolayer, in the case of direct
excitons in a Xene monolayer, $H_{0}$ and $Y_{0}$ are the Struve and Bessel
functions of the second kind, respectively, and $\rho _{0}$ is the
screening length. The screening length $\rho _{0}$ can be written as~\cite%
{Berkelbach2013} $\rho _{0}=(2\pi \chi _{2D})/(\kappa )$, where $\chi _{2D}$
is the 2D polarizability, which in turn is given by~\cite{Keldysh} $\chi
_{2D}=l\epsilon /4\pi $, where $\epsilon $ is the bulk dielectric constant
of the Xene monolayer.

Following Refs. \cite{Elliott, Gorkov,Lozovik,Shinada,Akimoto,Herold}, in
Eq. (\ref{eq:ehschro}) we introduce the coordinate of the center-of-mass $%
\mathbf{R}=\frac{m_{e}\mathbf{r}_{e}+m_{h}\mathbf{r}_{h}}{M}$ and the
relative motion coordinate $\mathbf{r}=\mathbf{r}_{e}-\mathbf{r}_{h},$ where the total
mass of the system is $M=m_{e}+m_{h}$, and consider the magnetic field
pointing in $z$-direction that is perpendicular to the monolayer where the
exciton is located, $\mathbf{B}=(0,0,B)$. After performing the standard
procedure for the coordinate transformation to the center-of-mass, the
Hamiltonian for Eq. (\ref{eq:ehschro}) becomes:

\begin{eqnarray}
H &=&-\frac{ 1}{2M}\frac{{\partial }^{2}}{\partial \bm{R}^{2}}-\frac{%
1}{2\mu }\frac{{\partial }^{2}}{\partial \bm{r}^{2}}+\frac{e^{2}}{%
8\mu }(\bm{B}\times \bm{R})^{2}+\frac{e^{2} \mu ^2}{8}\left( \frac{1}{m_e ^3}+\frac{1}{m_h ^3}\right) (\bm{B}\times \bm{r})^{2}-\frac{%
i e}{2M}(\bm{B}\times \bm{r})\cdot \frac{\partial }{\partial \bm{R}}
\notag \\
&&-\frac{i e}{2\mu }(\bm{B}\times \bm{R})\cdot \frac{\partial }{%
\partial \bm{r}}-\frac{i e\gamma }{2\mu }(\bm{B}\times \bm{r})\cdot
\frac{\partial }{\partial \bm{r}}+\frac{e^{2}\gamma }{4\mu }(\bm{B}%
\times \bm{R})\cdot (\bm{B}\times \bm{r})+V(\bm{r}),  \label{eq:HRr}
\end{eqnarray}%
where $\gamma =\frac{m_{h}-m_{e}}{m_{h}+m_{e}}$ and $\mu =\frac{m_{e}m_{h}}{%
m_{e}+m_{h}}$ is the reduced mass.
It is worth noting that (\ref{eq:HRr}) is written for the case when masses of the electron and hole are different: $m_{h}\neq m_{e}$.
In the case of the Xenes, the masses of electrons and holes are equal, $m_{h}=m_{e}=m$, and, therefore, $\gamma =0$ and $\mu =\frac{m}{2}$. The latter leads
to a significant simplification of the Hamiltonian (\ref{eq:HRr}). The terms $\frac{i e\gamma }{2\mu }(\bm{B}\times \bm{r})\cdot
\frac{\partial }{\partial \bm{r}} = 0$ and $\frac{e^{2}\gamma }{4\mu }(\bm{B}%
\times \bm{R})\cdot (\bm{B}\times \bm{r}) = 0$ when $\gamma =0$.

At the next step following Ref. \cite%
{Gorkov,Lozovik,Herold} we introduce an operator $\hat{\bm{P}}$, which
commutes with the Hamiltonian of Eq. (\ref{eq:ehschro}), and is defined as:
\begin{equation}
\hat{\bm{P}}=-i \nabla _{\bm{R}}-\frac{e}{2}(\bm{B}\times \bm{r}).
\label{eq:poperator}
\end{equation}%
Since $\hat{\bm{P}}$ commutes with the Hamiltonian (\ref{eq:HRr}),\ it has the
same eigenfunction as Eq. (\ref{eq:ehschro}). Therefore, one can write the
wave function for the exciton in the magnetic field as \cite{Gorkov, Lozovik}%
:
\begin{equation}
\psi (\bm{R},\bm{r})=e^{\left[ i\bm{R}\cdot (\bm{P}+\frac{e}{2%
}\bm{B}\times \bm{r})\right] }\Phi (\bm{r}-\tilde{\bm{\rho}}_0),  \label{eq:wavefun}
\end{equation}%
where we take
into account that $\gamma =0$ and $\tilde{\bm{\rho}}_0=\frac{1}{eB^2}(\bm{B}\times \bm{P})$. After substituting the
wave function $\psi (\bm{R},\bm{r})$ in Eq. (\ref{eq:HRr}), the Schr\"{o}dinger equation for the relative
motion of the electron and hole reads:

\begin{equation}
\left[ \frac{\bm{P}^2}{4 m} + \frac{e}{2m}\bm{P}\cdot (\bm{
B \times \bm{r}})-\frac{1}{m}\frac{{\partial }^{2}}{\partial \bm{r}%
^{2}}+\frac{e^{2}}{4m}(\bm{B}\times \bm{r})^{2}-\frac{ie}{m}(%
\bm{B}\times \bm{R})\cdot \frac{\partial }{\partial \bm{r}}+V(r)\right]
\Phi =\mathcal{E}\Phi.   \label{Relmotion}
\end{equation}%
Finally, the equation for a 2D electron-hole pair
with zero center-of-mass momentum reads \cite{MacDonald1986, Lozovik}:

\begin{equation}
\left[ -\frac{1}{m}\frac{{\partial }^{2}}{\partial \bm{r}^{2}}+\frac{%
e^{2}}{4 m}(\bm{B}\times \bm{r})^{2}+V(r)\right] \Phi (\bm{r})=\mathcal{E}\Phi (\bm{r}).  \label{eq:finschang}
\end{equation}%
After separating the angular variable, Eq. (\ref{eq:finschang}) can be rewritten as:

\begin{equation}
\left[ -\frac{1}{m}\frac{{\partial }^{2}}{\partial r^{2}}-\frac{1}{m}\frac{1}{r}\frac{{\partial }}{\partial r}+\frac{%
e^{2}}{4 m}(\bm{B}\times \bm{r})^{2}+V(r)\right] \Phi (r)=\mathcal{E}\Phi (r).  \label{eq:finsch}
\end{equation}%
Equation (\ref{eq:finsch}) describes the Mott--Wanner magnetoexciton in Rydberg states in Xenes. This equation has a long history of the solution in the
case of the electron-hole Coulomb interaction \citep{Elliott,
Gorkov,Shinada,Akimoto,Shinada1970,Lozovik1978,Herold,MacDonald1986,Stafford1990,Lozovik}%
. In earlier works, equation for the Mott--Wanner magnetoexcitons (\ref%
{eq:finsch}) has been previously solved by Wentzel-Krainers-Brillouin method
\cite{Akimoto}, by numerical integration \cite{Shinada1970}, using Pad\'{e}
approximation based on perturbation expansions about the weak- and
strong-field limits \cite{MacDonald1986,Stafford1990}. We cite these early
works, but the recent literature on the subject is not limited by them.

To find the binding energy of the direct magnetoexciton in Xenes, we use the
code implemented in Ref. \cite{Brunetti2018}. The code was modified in a way
that the Schr\"{o}dinger equation explicitly includes $\frac{e^{2}}{4 m}(%
\bm{B}\times \bm{r})^{2}$ term and can be solved for the effective mass $m$
that varies with the application of the external electric field. The latter allows
us to numerically solve Eq. (\ref{eq:finsch}), which includes the
contribution of electric and magnetic fields. In our numerical calculations, the maximum number of iterations for calculations of the binding energy is $10^{12}$, and increasing this number does not affect binding energies. The increase of the maximum number of iterations from $10^{10}$ to $10^{12}$ gives the change of the binding energy less than $10^{-3}$\%. The implementation of this code for TMDCs materials in Ref. \cite{Spiridonova} reproduce experimental binding energies reported in Refs. \citep{ Stier2018, Gor2019, Liu2019} within 7\%, and theoretical binding energies given in Refs. \cite{Kylanpaa, donckexc}, obtained using the stochastic variation and the path integral Monte-Carlo methods, respectively, within 1\%. The benefit of our approach is
that we can obtain the eigenfunctions and eigenenergies of the
magnetoexcitons for any value of the electric and magnetic fields. In our
calculations, we do not have a linear term (the Zeeman shift) \cite{Nash} since we consider the Rydberg $s$ states for which $\bm{B}\cdot \bm{L}=0
$ and we do not consider spin - magnetic field interaction, $\bm{S}\cdot %
\bm{B}=0$ \cite{Nash}. Therefore, the energy contribution from the electric and magnetic fields to the binding energy that we calculate is the
diamagnetic shift.

The diamagnetic coefficient has not been reported in the literature for magnetoexcitons in Xenes monolayers. However, DMCs were calculated and reported for magnetoexcitons in TMDCs monolayers. In contrast to our approach of calculating the diamagnetic coefficient, in Refs. \cite{Liu2019, Stier2018, Stier_2016} the authors
solve the Schr\"{o}dinger equation
without magnetic field and treat the magnetic field as the small
perturbation. In Ref. \cite{Gor2019} authors developed the model for revealing
exciton masses and dielectric properties of the monolayer semiconductors
with high magnetic fields by numerically solving the corresponding Schr\"{o}%
dinger equation with the RK potential, while in Ref. \cite{Liu2019}, the
corresponding Hamiltonian was diagonalized to find the DMCs.

\subsection{Indirect magnetoexcitons in Xenes heterostructure}
{\label{ssec:indirect}

Let us now consider the indirect magnetoexcitons formed by
electrons and holes located in two different Xene monolayers which are
separated by $N$ layers of hBN monolayers. Such magnetoexcitons
have a longer lifetime than the direct excitons due to longer recombination time. The electron and
hole via electromagnetic interaction $V(r)$, where $r$ is the distance between
the electron and hole in different layers, could form a bound state, i.e., a
magnetoexciton, now in 3D space. Therefore, to determine
the binding energy of the magnetoexciton one must solve a two-body problem in
restricted 3D space because the motion in $z$-direction is restricted. The relative separation $r$ between the
electron and hole can be written in cylindrical coordinates as:
 \begin{equation}
\bm{r}=\rho \hat{\bm{
\rho}}+D\hat{\bm{z}}.
\label{eq:cylcoord}
 \end{equation}
In Eq. (\ref{eq:cylcoord}), $\hat{\bm{\rho}}$ and $\hat{\bm{z}}$ are unit vectors. Writing $r$ in cylindrical coordinates allows us to treat the case of direct excitons in a Xene
monolayer and spatially indirect excitons in X-hBN-X heterostructures on
equal footing. If we set $D=0$, and Eq.~(\ref{eq:cylcoord}) becomes a purely
2D equation, with $\rho $ representing the separation between the electron
and hole sharing the same plane. Throughout this paper, we consider the
separation between two Xene monolayers in steps of calibrated thickness, $%
l_{\text{hBN}}=0.333~\text{nm}$, corresponding to the thickness of one h%
-BN monolayer. For spatially indirect excitons, the relative coordinate $r=%
\sqrt{\rho ^{2}+D^{2}}$, where $D=l+Nl_{\text{hBN}}$, $l$ is the Xene
monolayer thickness and $N$ is the number of hBN monolayers.

\bigskip
We perform calculations using both the RK and Coulomb potentials. This allows a better understanding of the importance of the screening effect in
X-hBN-X heterostructures. For indirect excitons, the expressions for the
interaction between the electron and hole in Eq. (\ref{eq:finsch}) can be
written as:}

\begin{equation}
V_{RK}(\sqrt{\rho ^{2}+D^{2}})=-\frac{\pi ke^{2}}{2\kappa \rho _{0}}\left[
H_{0}\left( \frac{\sqrt{\rho ^{2}+D^{2}}}{\rho _{0}}\right) -Y_{0}\left(
\frac{\sqrt{\rho ^{2}+D^{2}}}{\rho _{0}}\right) \right] ,  \label{eq:indkeld}
\end{equation}%
for the RK potential, and
\begin{equation}
V_C \left( \sqrt{\rho ^{2}+D^{2}}\right) =-\frac{ke^{2}}{\kappa \left( \sqrt{\rho
^{2}+D^{2}}\right) }  \label{eq:indcoul}
\end{equation}%
for the Coulomb potential. Equations~\eqref{eq:indkeld} and~\eqref{eq:indcoul}
describe the interaction between the electron and hole that are located in different parallel Xenes monolayers separated by a distance $D$. Therefore, one
can obtain the eigenfunctions and eigenenergies of magnetoexcitons by
solving Eq.~ (\ref{eq:finsch}) using the potential~(\ref{eq:rk}) for direct
magnetoexcitons, or using either potential~(\ref%
{eq:indkeld}) or~(\ref{eq:indcoul}) for indirect magnetoexcitons.

It is worth mentioning that the RK potential was originally formulated as an
explicitly 2D description of the Coulomb interaction.
Nevertheless, there have been recent attempts to apply the RK potential to
indirect excitons in van der Waals heterostructures of 2D
materials such as the TMDCs, phosphorene, and Xenes~\cite%
{Fogler2014,Berman2016,Berman2017,Berman2017a,Brunetti2018,BBKPRB2019}. The
logic behind considering the RK potential for indirect excitons
follows from two considerations: (i) the dielectric environment is still
inhomogeneous, just as in the case of the direct exciton --{} when the
interlayer separation $D$ is smaller than, or comparable to, the RK potential
screening length $\rho _{0}$ and the excitonic gyration radius $\sqrt{%
\langle r^{2}\rangle }$ - the electron-hole interaction potential must
account for both the Xene monolayers and the interlayer dielectric, and (ii)
as the interlayer separation $D$ becomes larger than $\rho _{0}$, the total
separation, $r=\sqrt{\rho ^{2}+D^{2}}$, between the electron and hole
necessarily becomes much larger than $\rho _{0}$, and therefore the RK
potential converges towards the Coulomb potential. Let us emphasize
that we are not claiming definitively that the RK potential provides the
most accurate description of the spatially indirect exciton. Therefore, we are comparing the results obtained using $V_C$ and $V_{RK}$ potentials.

\section{\label{sec:results} Results of calculations and discussion}

We report the energy contribution from the electric and magnetic fields to the binding energy of the magnetoexcitons in Rydberg states 1$s$, 2$s$, 3$s$, and 4$s$, in the freestanding silicene (FS Si), germanene (FS Ge), and stanene (FS Sn) (Fig. 1$a$), encapsulated silicene (Si type I, Si type II) \cite{LiWang2013} monolayers (Fig. 1$b$), and Xenes heterostructure (Fig. 1$c$) as a function of the external electric and magnetic fields perpendicular to the monolayer or heterostructure. The results of calculations of diamagnetic coefficients for excitons in the above mentioned systems and their dependence on the external electric field are presented as well. The external electric field, $E$, does not appear in the final Schr\"{o}%
dinger equation that we are solving. However, the dependence of the binding energies and DMCs on the external electric field enters the Schr\"{o}%
dinger equation through the effective masses of the electron and hole.

In our calculations, we vary the external magnetic field in the range from 0 T to 30 T in the increment of 1 T. We take a value of the external electric field, $E$, starting above the critical value $E_c$ in the increment of 0.1 V/{\AA} up to 2.7 V/{\AA} \cite{Drummond2012}. It is worth noting that at $E_c$ occurs a phase transition in freestanding monolayer Xenes from the semiconducting phase to the excitonic insulating phase \cite{BBKPRB2019} and $E_c$ is unique for each material.
Our calculations use input parameters given in Table \ref{table:parametersxenes}. Parameters of  the Xene monolayer depend on the substrate. For example, authors of Ref. \cite{LiWang2013} have found that there are nine possible Si-hBN stacking arrangements where interactions between Si and hBN significantly modify the properties of the Si: increase band gap and decrease Fermi velocity. Out of nine stacking arrangements given in \cite{LiWang2013}, one arrangement has the lowest band gap and Fermi velocity (Si type I), and one arrangement has the highest band gap and Fermi velocity (Si type II). Therefore, Si type I and type II parameters correspond to the upper and lower bounds on the exciton masses, respectively. 

\begin{table}[b]
\caption{Parameters for Xenes. Parameters used to calculate $m$ and the binding energies. FS refers to the freestanding monolayers (monolayer in a vacuum). $\kappa=\frac{\epsilon_1 + \epsilon_2}{2}$ is the average dielectric constant of encapsulating materials. 2$\Delta_{\text{so}}$ is the total gap between conduction and valence bands; $d_0$ is the buckling parameter; $v_F$ is the Fermi velocity; $l$ is the monolayer thickness; $\epsilon$ is the dielectric constant of the Xenes monolayer. $\chi_{2D}=\frac{l \epsilon}{4\pi}$ \cite{Keldysh} is the polarizability. }

\label{table:parametersxenes}
\centering
\begin{tabular}{P{2.5 cm}P{3cm}P{3cm}P{3cm}P{2.5cm}P{2.5cm}}

\hline \hline
 \multirow{1}{*}{} &
  \multicolumn{1}{c}{Si FS}  &
  \multicolumn{1}{c}{Ge FS} 
   &
  \multicolumn{1}{c}{Sn FS} &
  \multicolumn{1}{c}{Si (hBN, type I)}
  &
  \multicolumn{1}{c}{Si (hBN, type II)}
   \\ \cline{2-5}

  \hline
$\kappa$ & 1 & 1 & 1 & 4.89 & 4.89 \\

 2$\Delta_{\text{so}}$ (meV) & 1.9 (1.55)\cite{Matthesparam} (\cite{LiuFeng2011})& 33 (23.9) \cite{Matthesparam} (\cite{LiuFeng2011})& 101 (73.5) \cite{Matthesparam} (\cite{Liu2011})& 27 \cite{LiWang2013} & 38 \cite{LiWang2013}\\

 $d_0$ (\AA) & 0.46 \cite{Ni2012} & 0.676  \cite{Ni2012} & 0.85 \cite{Matthesparam} & 0.46 \cite{LiWang2013} & 0.46 \cite{LiWang2013} \\

 $v_F$($\times$10$^5$ m/s) & 6.5 (5.3) \cite{Matthesparam} (\cite{Matthes2013}) & 6.2 (5.2) \cite{Matthesparam} (\cite{Matthes2013})& 5.5 (4.8) \cite{Matthesparam} (\cite{Liu2011}) & 4.33 \cite{LiWang2013} & 5.06 \cite{LiWang2013} \\

 $\epsilon$ &  11.9 & 16 & 24 & 11.9 & 11.9\\

 $l$ (nm) & 0.4 \cite{Tao2015} & 0.45 & 0.5 & 0.333 \cite{LiWang2013} & 0.333 \cite{LiWang2013} \\

 $\chi_{2D}$(\AA) & 3.788 & 5.730 & 9.549 & 3.156 & 3.156\\
\hline \hline
\end{tabular}
\end{table}


\subsection{Contributions from external electric and magnetic fields to binding energies of magnetoexcitons}

\begin{figure}[t]
\begin{centering}
\includegraphics[width=16.0cm]{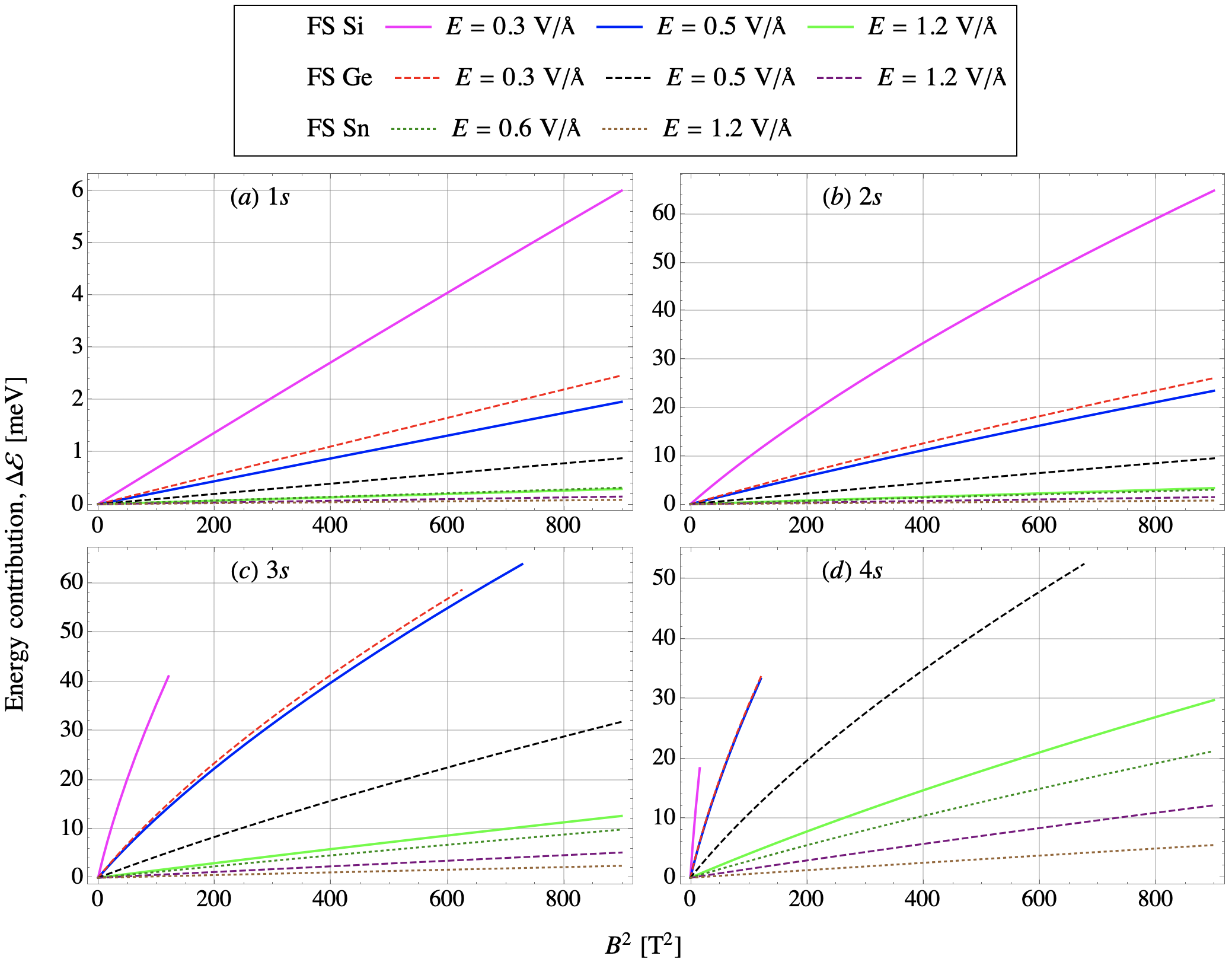}
\caption{The energy contribution to the binding energies of magnetoexcitons in FS monolayers in states 1$s$, 2$s$, 3$s$, and 4$s$ in ($a$), ($b$), ($c$), and ($d$), respectively, as a function of squared magnetic field. Data is plotted at three different values of the electric field for FS Si and FS Ge and at two values for FS Sn. The values of electric field are chosen in the following way: a value above $E_c$, another values at which significant energy contribution is present, and one representative value at which energy contribution converges to zero. The magnetic field, where the dissociation of magnetoexcitons in states 3$s$ and 4$s$ occurs, corresponds to the end of broken curve. }
\label{fs_states}
\end{centering}
\end{figure}

The energy contribution, $\Delta \mathcal{E}$, to the binding energy of the magnetoexcitons from the external electric and magnetic fields at the given electric field is calculating in the following way:

\begin{equation}
\Delta \mathcal{E} = \mathcal{E}_0 - \mathcal{E}(B,E). \label{eq:contr}
\end{equation}
In Eq. (\ref{eq:contr}) $\mathcal{E}_0$ is the binding energy of the exciton in the absence of the electric and magnetic fields, while $\mathcal{E}(B,E)$ is the binding energy of the magnetoexciton at some values of the external electric and magnetic fields. The results of our calculations of energy contribution, $\Delta\mathcal{E}$, for direct $A$ magnetoexcitons in freestanding Xenes monolayers are reported in Fig. \ref{fs_states}. In Figs. \ref{fs_states}$c$ and \ref{fs_states}$d$ that correspond to states 3$s$ and 4$s$, the end of broken curves indication dissociation of magnetoexcitons in these states. According to Fig. \ref{fs_states}, the energy contribution from the electric and magnetic fields to the binding energy goes from the highest to the smallest in the following order: FS Si, FS Ge, FS Sn. The energy contribution due to the varying electric and magnetic fields is negligible for magnetoexcitons in FS Sn in states 1$s$, 2$s$, and 3$s$ and only affects the 4$s$ state. $\Delta \mathcal{E}$ is significant for FS Si and FS Ge in 1$s$, 2$s$, 3$s$ and 4$s$ states when the electric field $E_c < E < 0.5$ V/{\AA} and magnetic field $B > 10$ T; for the states 3$s$ and 4$s$ the contribution from both electric and magnetic fields to the binding energies of magnetoexcitons almost doubles when $E_c < E < 1$ V/{\AA} and $B > 15$ T. When $B > 10$ T, the dependence of $\Delta \mathcal{E}$ on the electric field is very small for all states of the FS Si, FS Ge, and FS Sn. Therefore, the best range for the tunability of the energy contributions induced by the electric and magnetic fields for FS Si and Ge are $E_c < E < 1$ V/{\AA} (corresponds to the small reduced mass of excitons) and $B > 10$ T.

\begin{figure}[b]
\begin{tabular}{cc}

\textit{(a)} 1$s$ &  \textit{(b)} 2$s$ \\
\
  \includegraphics[width=80mm]{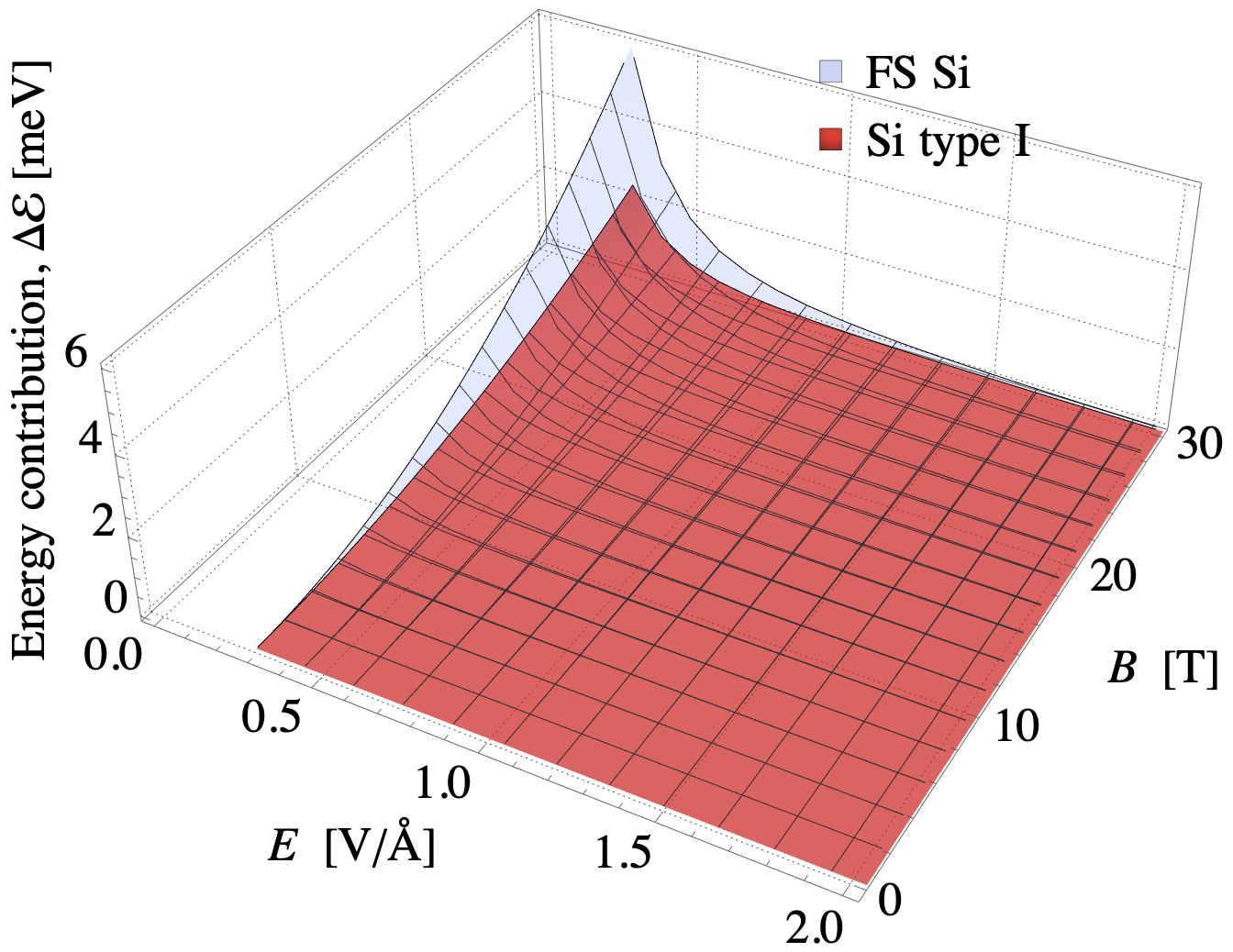} &   \includegraphics[width=80mm]{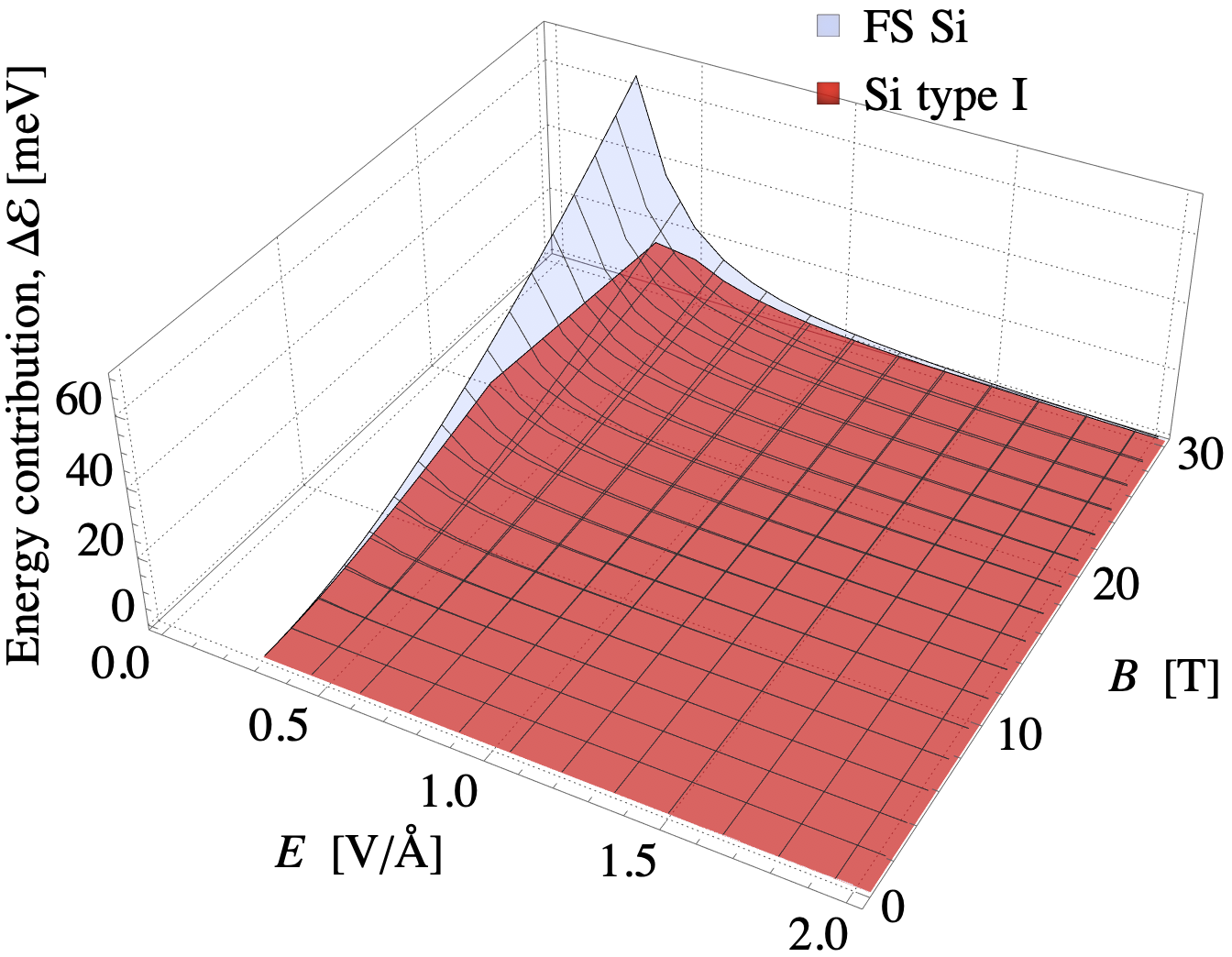} \\[6pt]

  \textit{(c)} 3$s$ & \textit{(d)} 4$s$ \\[6pt]
 \includegraphics[width=80mm]{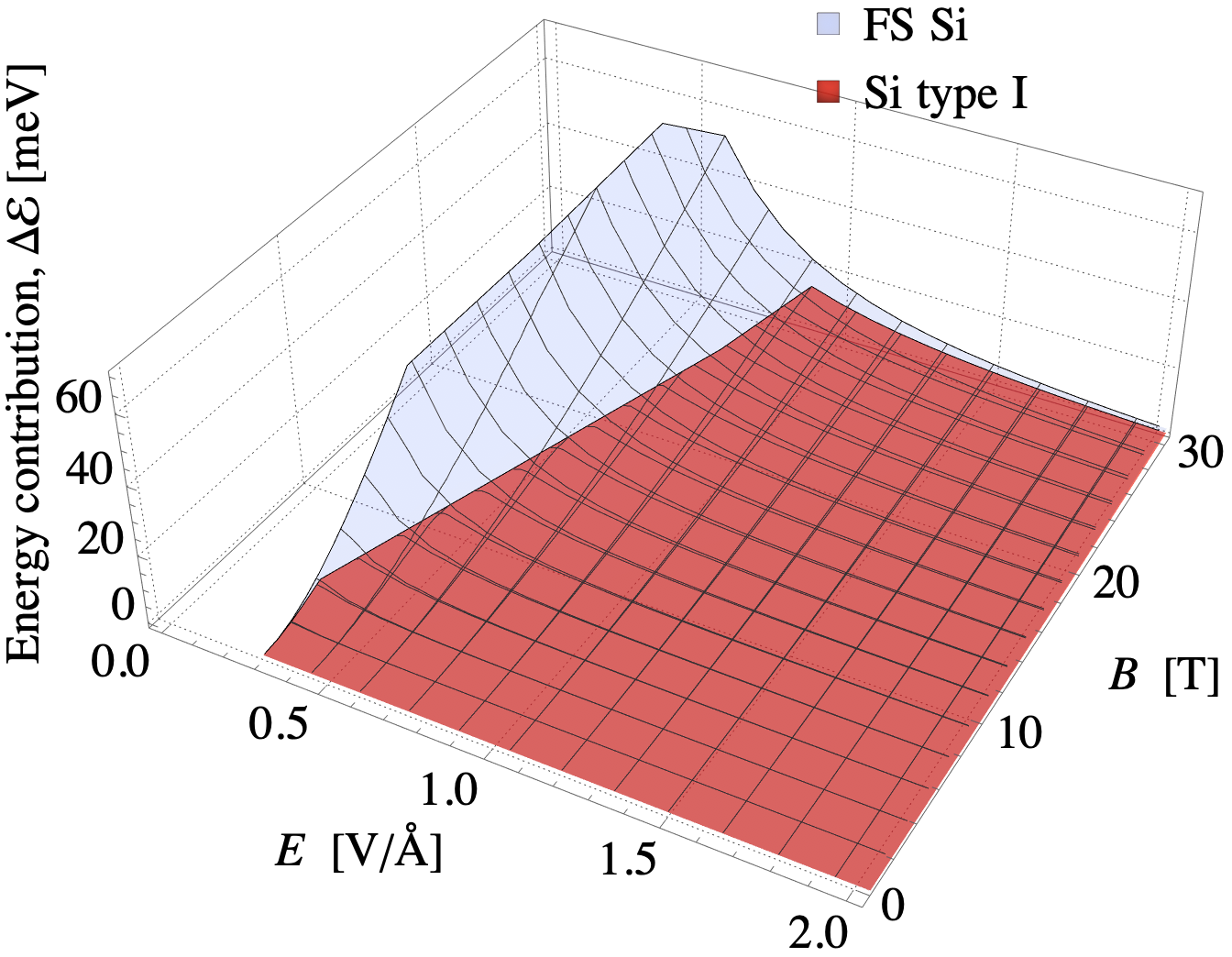} &   \includegraphics[width=80mm]{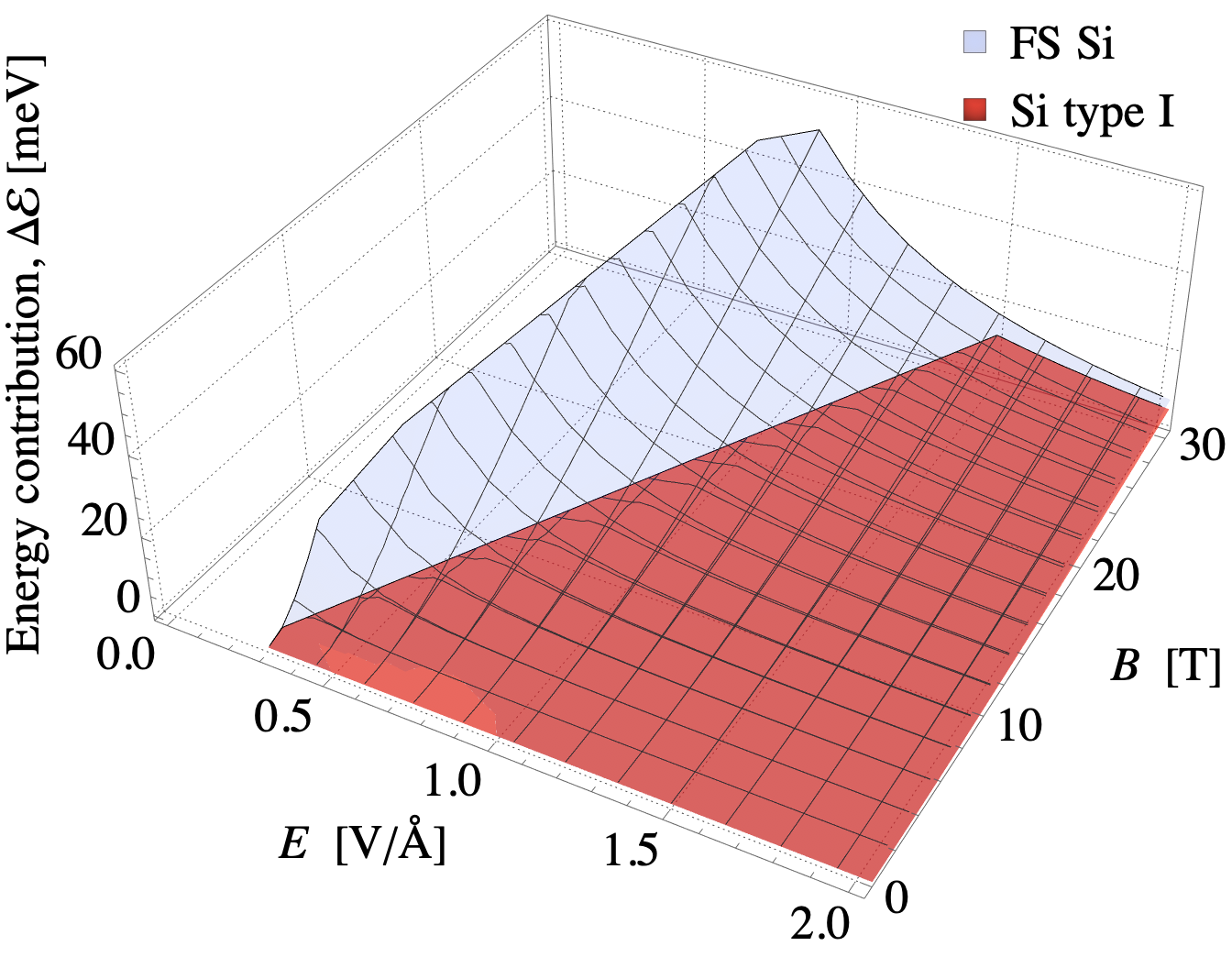} \\

\end{tabular}
\caption{The dependence of the energy contribution to the binding energy of direct magnetoexcitons on perpendicular electric and magnetic fields for freestanding Si and encapsulated Si type I monolayers. The data are plotted for states 1$s$, 2$s$, 3$s$, and 4$s$, and shown in ($a$), ($b$), ($c$), and ($d$), respectively. The broken surface edges correspond to the electric and magnetic fields at which the dissociation of magnetoexcitons occurs.} \label{si_states}
\end{figure}

Let us now compare the dependence of the energy contribution to the binding energy of direct magnetoexcitons on the electric and magnetic fields in a freestanding silicene and encapsulated Si type I. The results of calculations for 1$s$, 2$s$, 3$s$, and 4$s$ states are presented in Fig. \ref{si_states}. The direct magnetoexciton in freestanding Si monolayer has a smaller effective mass, $m$, than the magnetoexciton in silicene monolayers encapsulated by hBN. Also, the magnetoexciton in FS Si has the binding energy that is at least four times bigger than  binding energies of magnetoexcitons in Si type I.
As a result, magnetoexcitons in Si type monolayer at low values of $E$ dissociate at small values of the magnetic field, while magnetoexcitons in FS Si monolayers stay bound. It is worth mentioning that there are negligible differences in $\Delta \mathcal{E}$ for direct magnetoexcitons in encapsulated Si type I and Si type II for all values of the external electric and magnetic fields, which is not shown in Fig. \ref{si_states}. In addition, the energy contribution for magnetoexcitons in FS Si is always larger when the electric field $E < 0.6$ V/{\AA} and $B > 15$ T for the 1$s$ and 2$s$ states. The magnetoexcitons become unbound in 3$s$ and 4$s$ states in encapsulated Si type I monolayer with the increase of the magnetic field even for very small values of the electric field. Therefore, the difference between encapsulated and FS materials are due to substrate that gives different electron-hole masses which leads to different binding energies and energy contributions from the electric and magnetic fields to the binding energies of magnetoexcitons, especially for the 3$s$ and 4$s$ states. In addition, we note that magnetoexcitons in FS Ge and Si type I monolayers have similar contributions from the electric and magnetic fields to the binding energy when magnetoexcitons are bound. However, since magnetoexcitons in FS Ge monolayers have higher binding energies, magnetoexcitons in FS Ge stay bound while magnetoexcitons in Si type I monolayer  in states 3$s$ and 4$s$ at the magnetic field values we examine. This is a notable result that indicates that the contribution from the magnetic field to the binding energy of the magnetoexciton does not only depend on the effective mass, $m$, of electron and hole, but on material other parameters as well. The same kind of the picture can be observed if one compares the FS Ge and encapsulated Si type I and Si type II.

A careful examination of Table \ref{table:parametersxenes} indicates that there are different values of the Fermi velocity reported in literature \cite{Matthesparam, Matthes2013, Liu2011} for FS silicene, germanene and stanene. We examined how Fermi velocity affects $m$ and the contribution from the electric and magnetic fields to the binding energy of the states 1$s$, 2$s$, 3$s$, and 4$s$. We found effective mass of magnetoexcitons in FS Si monolayer using $v_F =6.5\times 10^5$ m/s \cite{Matthesparam} and  $v_F =5.3\times 10^5$ m/s \cite{Matthes2013} and calculated the binding energies of the magnetoexciton. We examine the dependence of the binding energy on $v_{F}$. Since it appears in the denominator and is squared, changing $v_{F}$ leads to the higher energy change. By calculating effective mass with different values of the band gap, the change in effective mass is  smaller, and, therefore, the change in the binding energy is also smaller. Using a smaller value of $v_F$ leads to a higher mass since $v_F$ appears in the denominator in Eq. (\ref{eq:effmassEz}).
As expected, results with a higher $v_F$ have a higher energy contribution. While the difference between two data sets is relatively small in the state 1$s$, in the states 2$s$, 3$s$, and 4$s$ the difference between two data sets is significant. This indicates two things. First, the exciton mass in Xenes is sensitive to the parameters of the material. Second, keeping all parameters the same except for $v_F$ indicates that $\Delta \mathcal{E}$ has a strong dependence on effective mass, $m$. Therefore, we can conclude that the binding energy of the magnetoexciton strongly depends on all material parameters given in Table \ref{table:parametersxenes}.

\begin{figure}
\begin{centering}
\includegraphics[width=19.0cm]{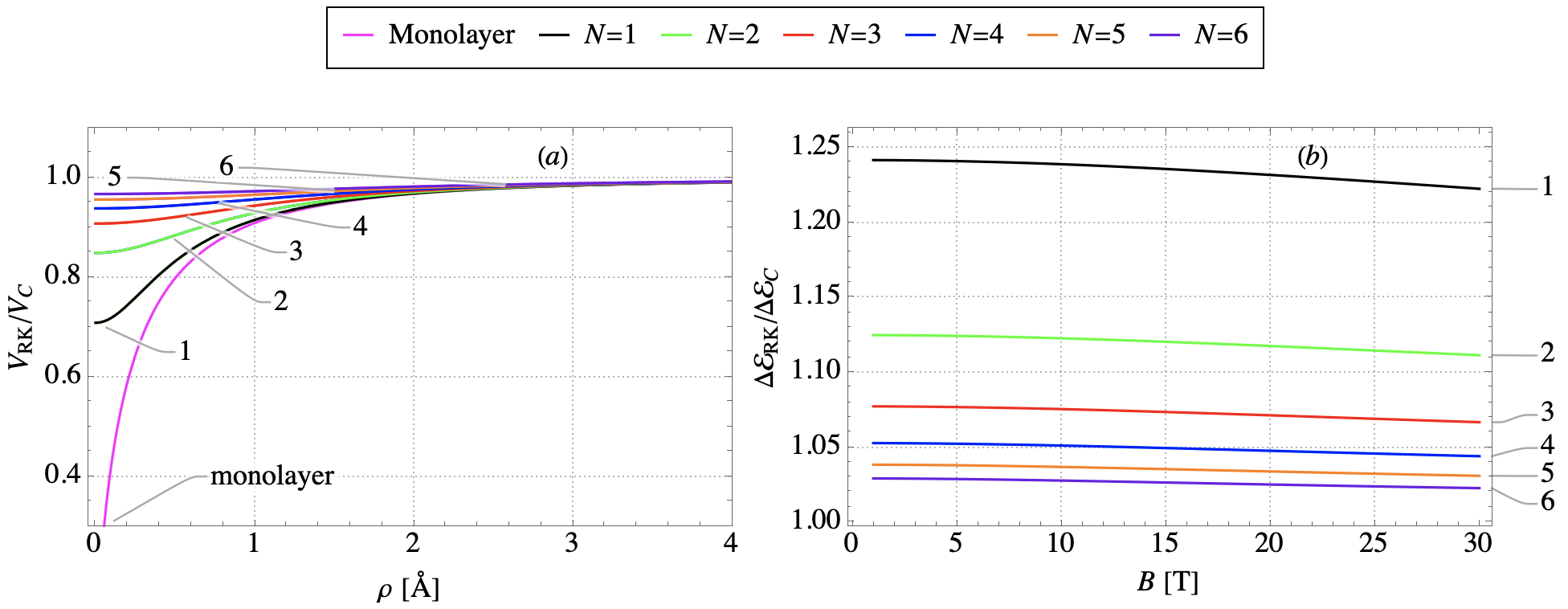}
\caption{ ($a$) The ratio of the RK to Coulomb potentials for electron-hole interaction in encapsulated Si type I monolayer and Xenes double-layer heterostructure as a function of the electron-hole separation. In case of the monolayer the $V_{RK}$ potential converges to the $V_{C}$ potential as the distance between the electron and hole increases. In the case of the Xenes double-layer heterostructure the $V_{RK}$ converges to $V_{C}$ as the distance between the projected electron and hole and the number of hBN layers, $N$, that separate two monolayer increase. ($b$) The ratio of the energy contributions to the binding energy of magnetoexcitons obtained using $V_{RK}$ and $V_{C}$ is plotted as a function of the magnetic field for varying number of hBN layers. Calculations are performed for magnetoexcitons in the Si type I monolayer for the state 1$s$.}
\label{pot_dif}
\end{centering}
\end{figure}

\begin{figure}
\begin{tabular}{cc}

\textit{(a)} $V_{RK}$ &  \textit{(b)} $V_C$\\
\
  \includegraphics[width=90mm]{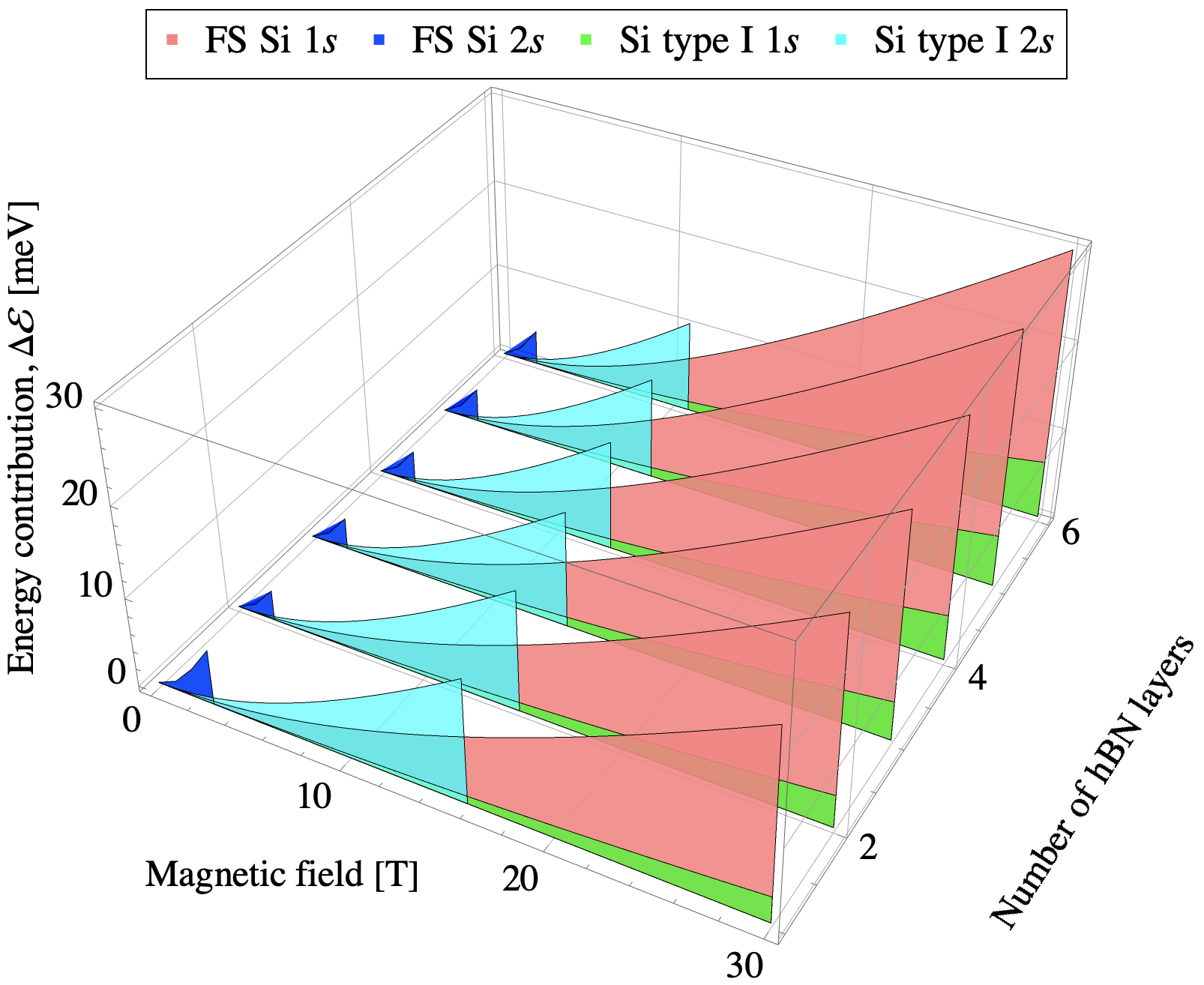} &   \includegraphics[width=90mm]{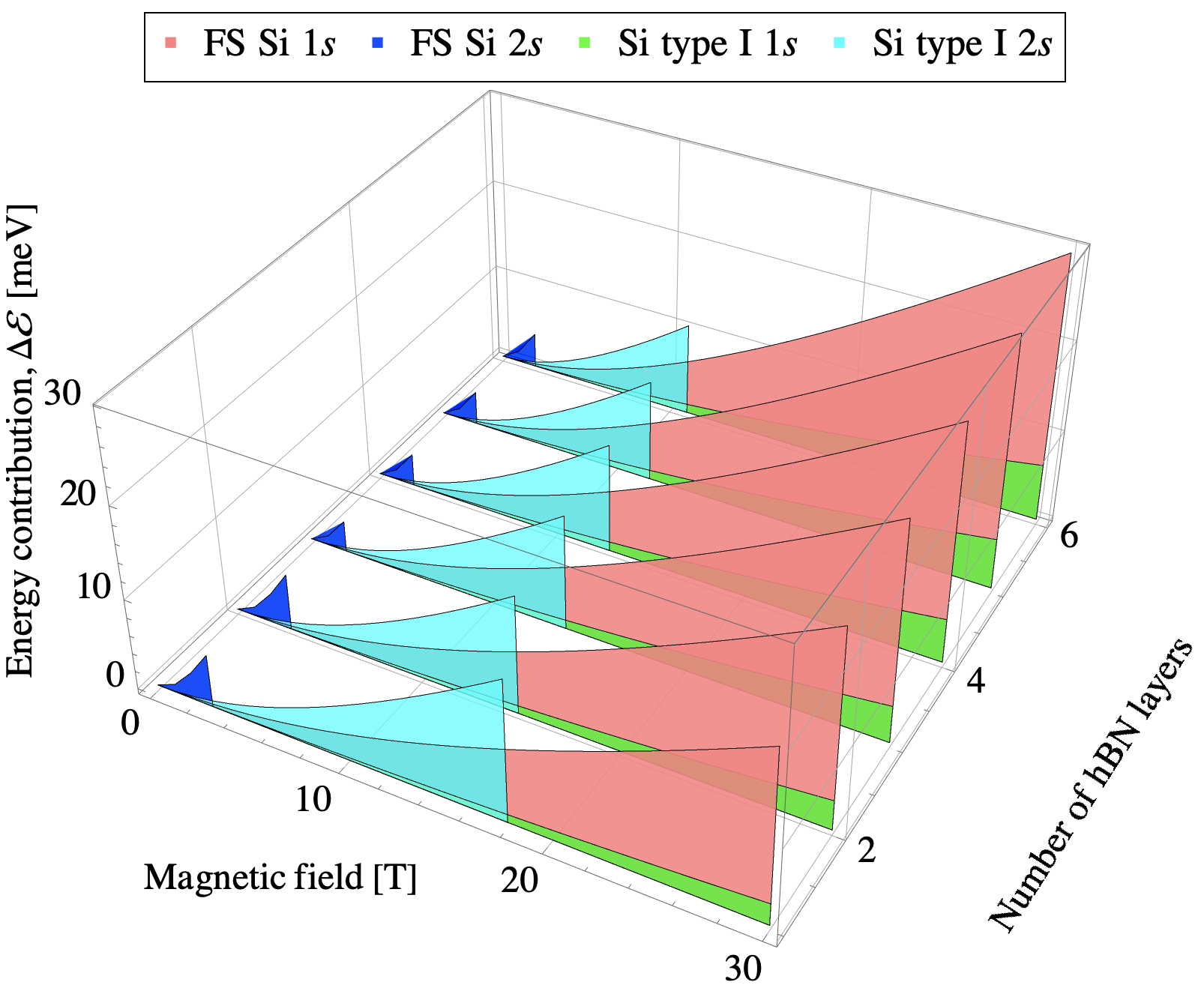} \\[6pt]

\end{tabular}
\caption{ The dependence of the energy contributions for indirect magnetoexcitons in FS Si and Si type I on the magnetic field for varying number of separating hBN monolayers. Calculations are performed using the RK ($a$) and Coulomb ($b$) potentials at $E=0.3$ V/{\AA}. The energy contribution for indirect magnetoexcitons in FS Si and Si type I double-layer heterostructures is plotted as a function of the magnetic field at the varying number of hBN layers. The surface edge tips for the 2$s$ state correspond to the magnetic field where the dissociation of the magnetoexciton occurs.}  \label{si_heter_contrib}
\end{figure}

Let us now consider the energy contribution to the magnetoexcitons binding energy from the electric and magnetic fields in X-hBN-X. In a heterostructure, Xenes monolayers are separated by $N$ ($N$=1, 2, 3, 4, 5, 6) hBN monolayers. The binding energies are calculated by using both the RK and Coulomb potentials. The latter allows to demonstrate the importance of the screening in Xenes heterostructures.
A comparison of the RK and Coulomb potentials for an electron-hole pair in the Xene heterostructure Si-hBN-Si is shown in Fig. \ref{pot_dif}$a$. On one hand, according to Fig. \ref{pot_dif}$a$, the RK potential is weaker than the Coulomb potential at small projections, $\rho$, of the electron-hole distance on the monolayer
plane, and their difference significantly decreases with the increase of the number of hBN layers. On the other hand, both potentials converge to each other as $\rho$ and the number of hBN layers increase. From the known asymptotic properties of the Struve and Bessel functions \cite{Abramowitz, Ryzhik}, it is easy to show that $\displaystyle{\lim_{\rho \rightarrow 0}\frac{V_{RK}}{V_{C}}=\frac{\pi D}{2\rho _{0}}\left[H_{0}(\frac{D}{_{\rho _{0}}})-Y_{0}(\frac{D}{_{\rho _{0}}})\right]}$. In Fig. \ref{pot_dif}$a$, the ratio $V_{RK}/V_{C}$ for the Si type I monolayer as a function of the electron hole separation, $r_{eh}$, is presented. In this case $\displaystyle{\lim_{r_{eh}\rightarrow 0}\frac{V_{RK}}{V_{C}}=0}$. The results of calculations of the ratio $ \Delta\mathcal{E}_{RK}/\Delta\mathcal{E}_{C}$ for the SI type I monolayer is presented in Fig. \ref{pot_dif}$b$ and show that the Rytova-Keldysh potential gives much more contribution to the binding energies of the magnetoexciton than the Coulomb potential for all range of the magnetic field and when the number of hBN layers is less than four. As the number of hBN layers increases, the binding energies calculated with both potentials converge.

In calculations of the contribution of the electric and magnetic fields to the binding energy of magnetoexcitons in heterostructures, we focus on FS Si, Si types I and II to demonstrate the importance of using physically accurate parameters. Also, we focus on Si since there are no parameters available in literature for excitons in other encapsulated Xenes monolayers. The choice is related to the parameters of these materials listed in Table \ref{table:parametersxenes}. From Table \ref{table:parametersxenes}, it is clear that the parameters of FS Si differ from those for Si types I and II. We report results for indirect magnetoexcitons in Si FS and Si types I. In other words, to describe magnetoexcitons in FS Si for the direct magnetoexcitons we use parameters of FS Si with $\kappa =1$, and for indirect magnetoexcitons we use parameters of FS Si with $\kappa =4.89$. Therefore, we do not take into account how parameters of the FS monolayer change when it is deposited on hBN. We did so to demonstrate the importance of using physically accurate parameters of materials since it affects
the binding energies. The results for indirect magnetoexcitons in the states 1$s$ and 2$s$ in FS Si and Si type I are reported in Fig. \ref{si_heter_contrib}. The calculations are performed for $E=0.3$ V/{\AA}  using the RK and Coulomb potentials. In contrast to direct magnetoexcitons, in the heterostructure indirect magnetoexcitons in Si types I and II have higher binding energies than magnetoexcitons in FS Si. However, indirect magnetoexcitons in FS Si have a much higher energy contribution from the magnetic field. In overall, we have demonstrated that the energy contribution of magnetoexcitons in the monolayer and the double-layer heterostructure can be tuned by electric and magnetic fields.

In our discussion, we examine $A$ excitons because they have higher binding energies due to larger reduced masses. The results for $B$ excitons are very close to the results of $A$ excitons because above $E=0.3$ V/{\AA} there is a small difference between the masses of $A$ and $B$ excitons in the materials we examined. The energy contribution $\Delta \mathcal{E}$ to the binding energies of excitons is bigger for the smaller reduced mass of excitons. Because $A$ excitons have bigger reduced mass, one can consider the results of our calculations as the lower limit for the energy contribution. When two Xenes monolayers separated by hBN layers are brought together the conduction band minimum and the valence band maximum reside in two different layers forming indirect (interlayer) exciton. There are possible different stacking orders: AA and AB as in TMDCs \citep{Pflugradt2014, Yarmohammad2017, Zhang2020, Schneider2019}. The $A$ exciton, for which mass is defined by large gap, has the bigger effective masses, $m_{A}$, of the electron and hole than the $B$ exciton formed by electron and holes with masses $m_{B}$: $m_{A}>m_{B}$. Thus, the reduced mass of $A$ excitons $\mu_{A}$ is bigger than the reduced mass of $B$ excitons:  $\mu_{A} > \mu_{B}$. In case of AA stacking, the bright interlayer exciton is formed by the hole from the valence band in $K$ valley of Layer 1 and the electron from the conduction band in $K$ valley of Layer 2 \cite{Zhang2020}. So masses of the electron and hole are equal. In case of AB stacking, because the lower layer is 180$^{\circ}$ in plane rotation of the upper layer \cite{Jones2014}, band spins of $K$ valley of Layer 2 are flipped compared to corresponding band spins of $K$ valley of Layer 1. As a result, the bright interlayer exciton can be formed by the hole with $m_{A}$ in $K$ valley of the Layer 1 and the electron in $K$ valley of the Layer 2 with mass $m_{B}$ \citep{Arora2017,Horng2018,Gerber2019,Lorchat2021}. This interlayer exciton has the reduced mass $m_{A} m_{B}/(m_{A}+m_{B})$. Thus, $m_{B}/2 < m_{A} m_{B}/(m_{A}+m_{B}) <m_{A}/2$. So, when the electron and hole reside in separate layers and have different masses, the binding energy of this interlayer $K/K$ valleys exciton falls between binding energies of interlayer excitons formed by electron and hole with equal masses. Because the energy contribution due to the magnetic field is smaller for the bigger reduced mass, our calculations for $A$ magnetoexcitons provide the lower limit for $\Delta \mathcal{E}$. Therefore, results for direct and indirect $B$ magnetoexcitons and indirect magnetoexcitons with the reduced mass $m_{A} m_{B}/(m_{A}+m_{B})$ always give higher $\Delta \mathcal{E}$. In our approach, the results of calculations for $\Delta \mathcal{E}$ are only affected by the numerical value of the reduced mass, but all qualitative conclusions remain the same for all types of magnetoexcitons.

\subsection{\protect\bigskip Diamagnetic shifts}

The application
of external electric or magnetic fields can give valuable information
about the exciton. In particular, an external magnetic field has two
effects on the spectroscopically observed energy levels: the spin splitting of levels, which is linear with respect to the
applied field, and the diamagnetic shift - an increase
in energy of levels with the magnetic field. The diamagnetic shift of an exciton in 2D materials has
been studied by many of authors \citep{DonckDM2018,Liu2019, Stier_2016,Aivazian,Gor2019,Macneill,Plechinger,Stier2018,Luckert,donckexc, Choi2015,Walck,Han2018,Zipfel, Spiridonova}. So far diamagnetic shifts have not been considered in Xenes monolayers or double-layer heterostructures. It is our goal to
extend this work to Xenes. In this paper, we apply the analysis used to describe diamagnetic shifts in quantum dots and TMDCs. Here we present the framework for Zeeman and diamagnetic shifts for magnetoexcitons in Xene monolayers and double-layer heterostructures. According to Ref. \cite{Walck}, when the contribution to the binding energy from the magnetic field is small compared to the binding energy, the magnetoexciton binding
energy, $\mathcal{E}(B)$, for $A$ and $B$ excitons can be expanded in Taylor series
as:
\begin{equation}
\mathcal{E}(B)=\mathcal{E}_{0}+\gamma _{1}B+\gamma _{2}B^{2}+... \label{eq:taylorser}
\end{equation}%
In Eq. (\ref{eq:taylorser})
the terms $\gamma _{1}B$ and $\gamma _{2}B^{2}$ are
identified as the valley Zeeman and diamagnetic shifts, respectively.
The expansion of $\mathcal{E}(B)$ in terms of $B$ for the magnetoexciton energy given in Eq. (\ref{eq:taylorser}) is
applicable when $\mathcal{E}_{0}>|\mathcal{E}(B)-\mathcal{E}_{0}|$. This
expansion is no longer applicable, when $\mathcal{E}_{0}\sim |\mathcal{E}(B)-\mathcal{E}_{0}|$, and needs the consideration of the
next terms in the Taylor series. The valley Zeeman shift contribution to the binding energy is identified as $%
\gamma _{1}B=-\mu _{B}gB$, where $g$ and $\mu _{B}$ are $g$-factor and Bohr
magneton, respectively \cite{Aivazian,Macneill}. In Refs. \cite{Liu2019, Stier2018, Stier_2016, Gor2019, Plechinger}
the valley Zeeman shift for the magnetoexciton in the monolayer is defined as the energy difference between magnetoexcitons
located at Dirac points $K$ and $K^{\prime }$, $-g\mu _{B}B=\mathcal{E}(K)-\mathcal{E}(K^{\prime
})$ \cite{Liu2019, Stier2018, Stier_2016, Gor2019, Plechinger}. The diamagnetic shift contribution to the
binding energy is identified as $\gamma _{2}B^{2}=\frac{e^{2}B^{2}}{8\mu%
} \langle r^{2}\rangle $ \cite{Stier2018,Stier_2016}, where $\mu $ is the exciton
reduced mass, $e$ is the charge of the electron, and $\langle r^{2}\rangle $
is the expectation value of $r^{2}$ over the exciton envelope wave function. The factor
$\gamma _{2}$ is called the diamagnetic coefficient, and we denote it as $\sigma $. In the case of Xenes the reduced mass is $\mu = m/2$ and the diamagnetic shift
contribution to the binding energy is $\frac{e^{2}B^{2}}{4m%
}\langle r^{2}\rangle $. Experimentally the energy contribution from the diamagnetic shift is
defined as the average transition energy of each magnetoexciton state between
points $K$ and $K^{\prime }$, $\sigma B^{2}=\frac{\mathcal{E}(K)+\mathcal{E}(K^{\prime })}{2}$
\cite{Liu2019,Stier2018,Stier_2016,Chen2019,Macneill}. In TMDCs DMCs are used to determine exciton masses, radius, and dielectric properties of materials. For similar purposes, $\sigma$ can be used for excitons in Xene monolayers.

\begin{figure}[h]
\centering
\includegraphics[scale=0.6]{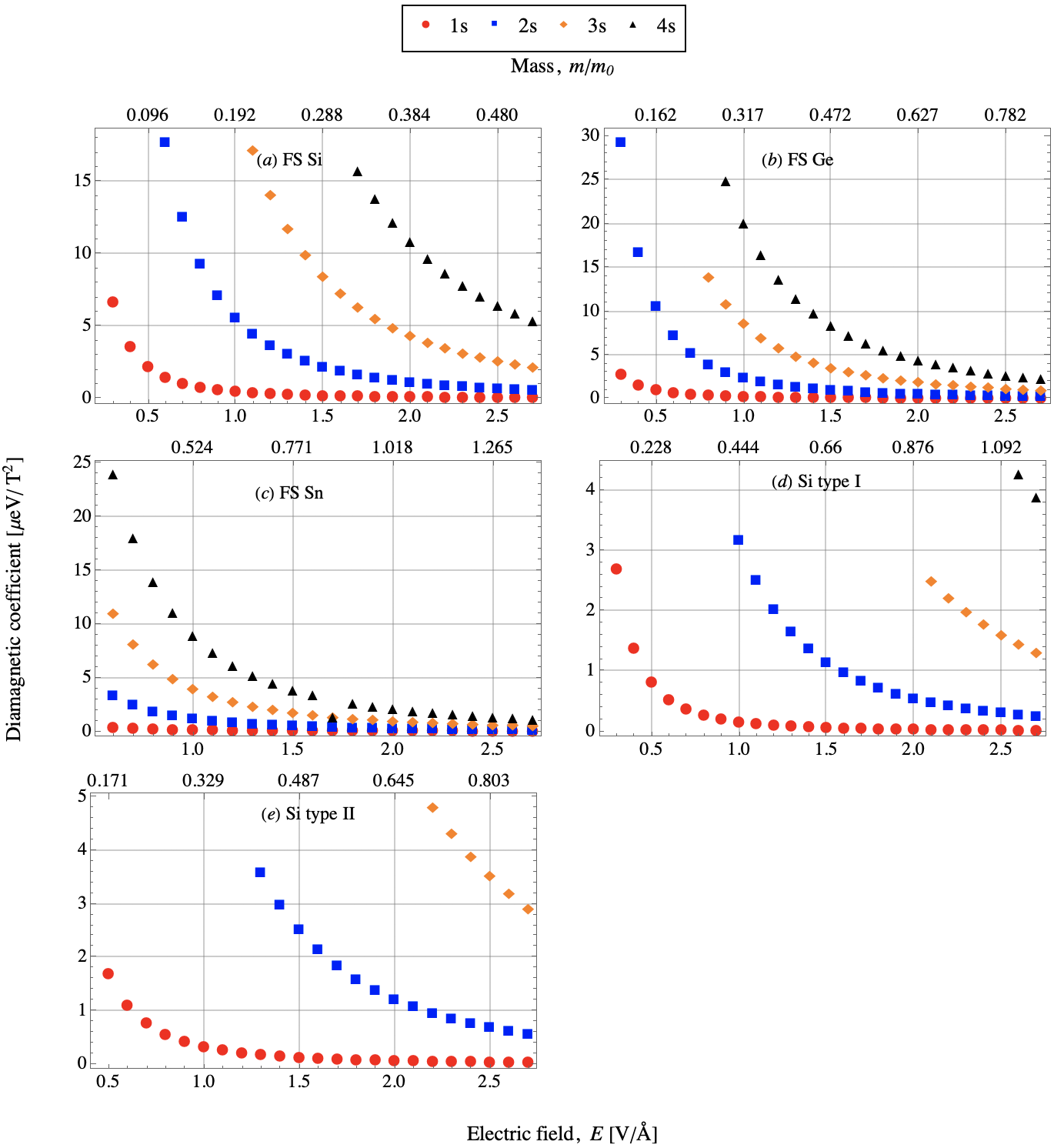}
\caption{The diamagnetic coefficients as a function of the perpendicular electric field for each material in states 1$s$, 2$s$, 3$s$, and 4$s$. At the bottom of each graph $E$ scale is given, and at the top of each graph $m/m_0$ scale is given.} \label{diamcoefmat}
\end{figure}

We have calculated the diamagnetic coefficients for $A$ magnetoexcitons in monolayers of FS Si, FS Ge, FS Sn, Si types I and II for the Rydberg states, 1$s$, 2$s$, 3$s$, and 4$s$. The DMCs for the direct $A$ magnetoexcitons in the above materials are reported in Fig. \ref{diamcoefmat}. The plotted data takes into account the critical value of $E$ below which the material behaves like excitonic insulator \citep{Knox,BBKPRB2019}. In Fig. \ref{diamcoefmat} at the bottom and the top of each graph scales for the electric field, $E$, and the corresponding effective mass, $m/m_{0}$, respectively, are given.
In contrast to $\sigma$ for magnetoexcitons in TMDCs monolayers reported only for 1$s$ and 2$s$ states in Ref. \cite{Spiridonova}, for magnetoexcitons in Xene monolayers $\sigma$ can be calculated for states 1$s$, 2$s$, 3$s$, and 4$s$.
At smaller values of the external electric field $E$,
the dependence of $\Delta \mathcal{E}$ on $B^2$ is not linear for magnetoexcotons in states 2s, 3s, and 4s. For direct magnetoexcitons at smaller values of $E$ higher order terms in Taylor expansion, Eq. (\ref{eq:taylorser}), need to be considered.  As can be seen from Fig. \ref{diamcoefmat}, in the states 2$s$, 3$s$, and 4$s$ magnetoexcitons have diamagnetic shifts
at electric field $E > 0.6$ V/{\AA}, while the magnetoexciton in FS Sn has $\sigma$ for all states when $E > E_{c}$.

As can be seen from Figs. \ref{diamcoefmat}$a$ and \ref{diamcoefmat}$b$, at the smaller value of $E$ in FS Si and FS Ge, respectively, parameters give lower reduced mass which leads to higher energy contribution and higher DMCs. According to Fig. \ref{diamcoefmat}, as $m$ increases the DMCs of direct magnetoexcitons in the examined states and materials decrease and approach 0. The condition $\mathcal{E}_{0}\sim |\mathcal{E}(B)-\mathcal{E}_{0}|$ exists for magnetoexcitons in some materials at low values of  $E$ in states 2$s$, 3$s$, and 4$s$, and we cannot apply Eq. (\ref{eq:taylorser}) to extract $\sigma$ from our data. 
The condition $\mathcal{E}_{0}\sim |\mathcal{E}(B)-\mathcal{E}_{0}|$ is prominent in Si types I and II monolayers because magnetoexcitons in Si types I and II monolayers in states 2$s$, 3$s$, and 4$s$ have the binding energies four times smaller than excitons in corresponding states in monolayers of FS materials. Also as it can be seen from Fig. \ref{diamcoefmat}$e$, the magnetoexcitons in Si type II in the state 4$s$ has no diamagnetic shift. Note, that magnetoexcitons in Si types I and II monolayers and double-layer heterostructures have the same energy contribution to the binding energy and DMCs at the same values of $m$. However, the magnetoexcitons in Si types I and II have different critical values of $E$ and have different values of $m$ at the same value of $E$. On one hand, direct magnetoexcitons in Si types I and II monolayers at small values of $E$ quickly dissociate when the perpendicular magnetic field is applied to the monolayer. For example, the magnetoexciton in Si type I (type II) at $E=0.5$ V/{\AA} in state 3$s$ dissociates at 11 T (7 T) and in state 4$s$ at 5 T (3 T). On the other hand, direct magnetoexcitons in monolayers of the FS Xenes in states 2$s$, 3$s$, and 4$s$ stay bound at higher values of the magnetic field and satisfy the condition, $\mathcal{E}_{0}>|\mathcal{E}(B)-\mathcal{E}_{0}|$, even at small values of electric field, $E$, that correspond to the small mass $m$ and allow us to extract DMCs.

\begin{figure}[t]
\begin{tabular}{cc}

\textit{(a)} 1$s$ $V_{RK}$ &  \textit{(b)} 1$s$ $V_C$\\
\
  \includegraphics[width=90mm]{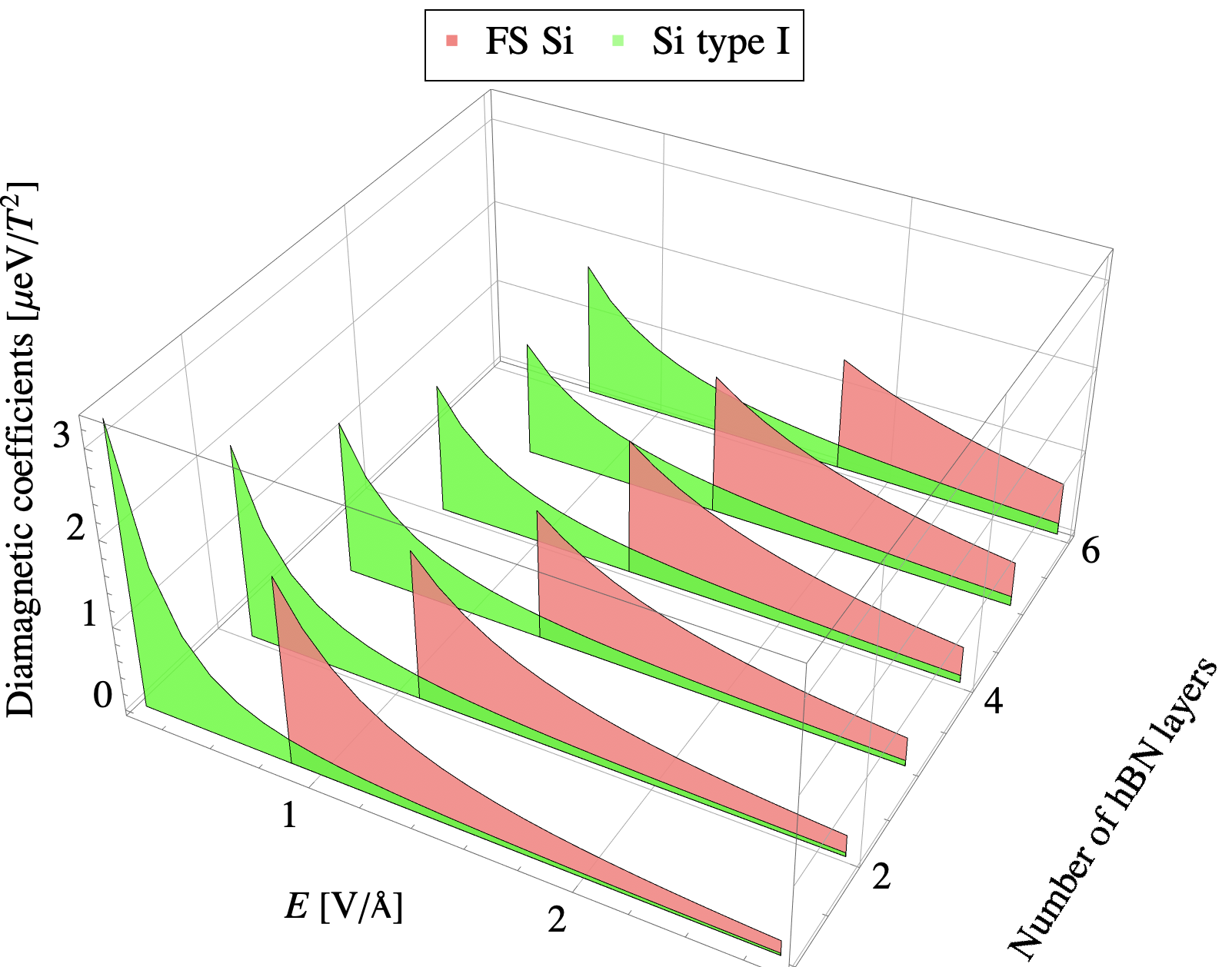} &   \includegraphics[width=90mm]{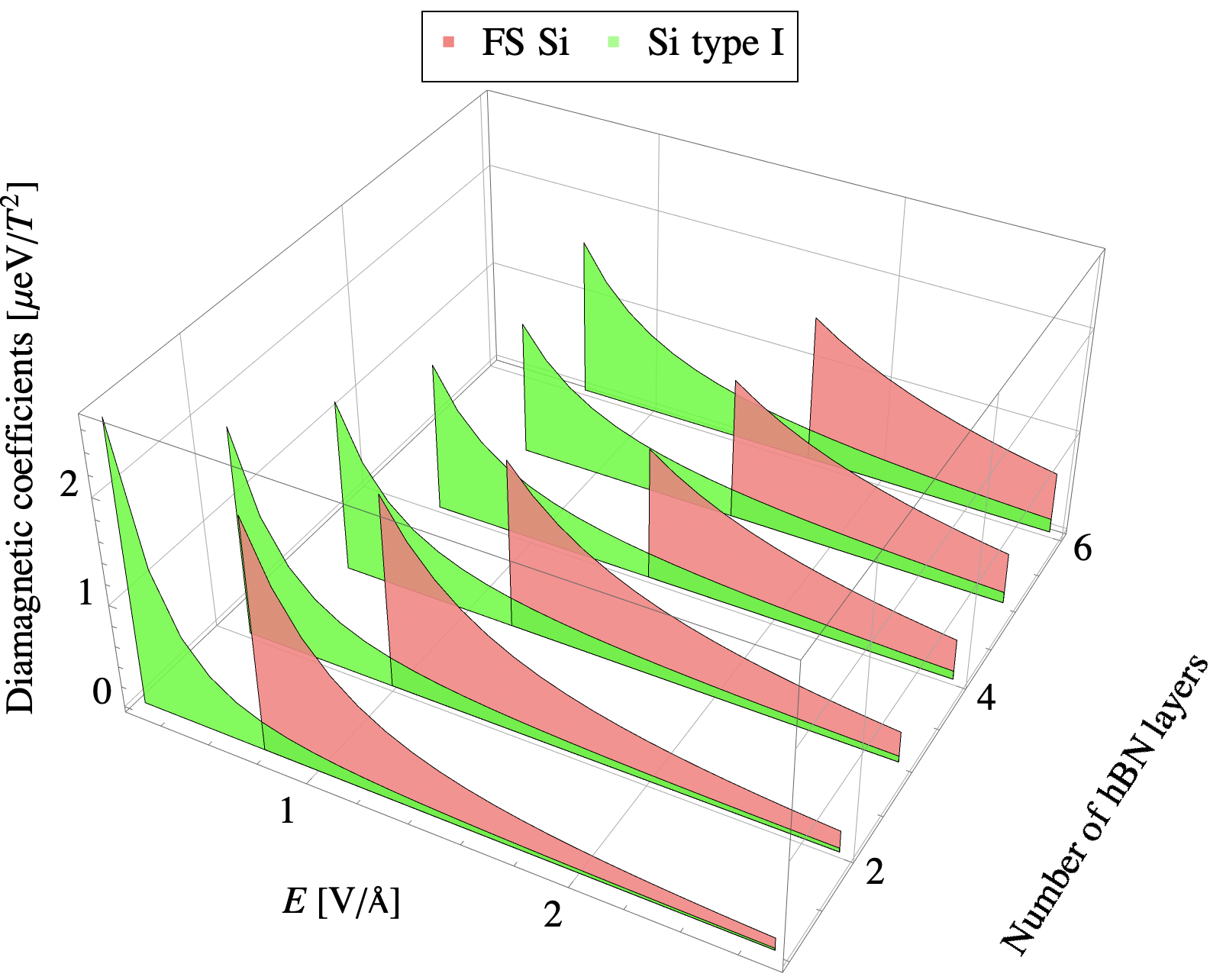} \\[6pt]

\end{tabular}
\caption{ Diamagnetic coefficients as a function of the perpendicular electric field at varying number of hBN layers that separate Xenes monolayers. $(a)$ Comparison of the DMCs of the magnetoexcitons in FS Si and Si type I when the Schr\"{o}dinger equation was solved with the Rytova-Keldysh potential. $(b)$ Comparison of the DMCs of the magnetoexcitons in FS Si and Si type I when the Schr\"{o}dinger equation was solved with the Coulomb potential. The surface edge tips correspond to lower values of the electric field above which DMCs can be extracted since there is linear dependence between $\Delta \mathcal{E}$ and $B^2$.}
\label{diam_coef_si_vrk_vc}
\end{figure}

For indirect $A$ magnetoexcitons in the double-layer heterostructure from the calculated binding energies with both the Rytova-Keldysh and Coulomb potentials, we extracted the DMCs of the magnetoexcitons for FS silicene and Si type I. The results for DMCs of indirect magnetoexcitons in the double-layer heterostructure of Xenes
are shown in Fig. \ref{diam_coef_si_vrk_vc}. The comparison of DMCs calculated from binding energies obtained with the RK and Coulomb potentials is shown only for indirect magnetoexcitons in FS Si and Si type I. Our calculations show that magnetoexcitons in other Xenes double-layer heterostructures have similar quantitative and qualitaive behavior. As the electric field increases (the corresponding effective masse of the electron and holes increases) and the number of hBN layers increases the DMCs decrease and approach zero. The DMCs for indirect magnetoexcitons calculated with $V_{RK}$ and $V_C$ potentials converge with the increase of the electric field, as can be seen from Fig. \ref{diam_coef_si_vrk_vc}. However, the DMCs obtained by using the RK potential are higher. Therefore, the choice of interaction potential is important when a few hBN layers separate the Xene monolayers. Another important effect of the number of hBN layers is that as the distance between Xene monolayers increases the energy contribution from the magnetic field to the binding energy does not have linear dependence on $B^2$. More specifically, as the
 distance between Xene monolayers increases the lowest value of $E$ at which linear dependence of $\Delta \mathcal{E}$ on $B^2$ exists increases.

\begin{figure}[t]
\centering
\includegraphics[scale=0.5]{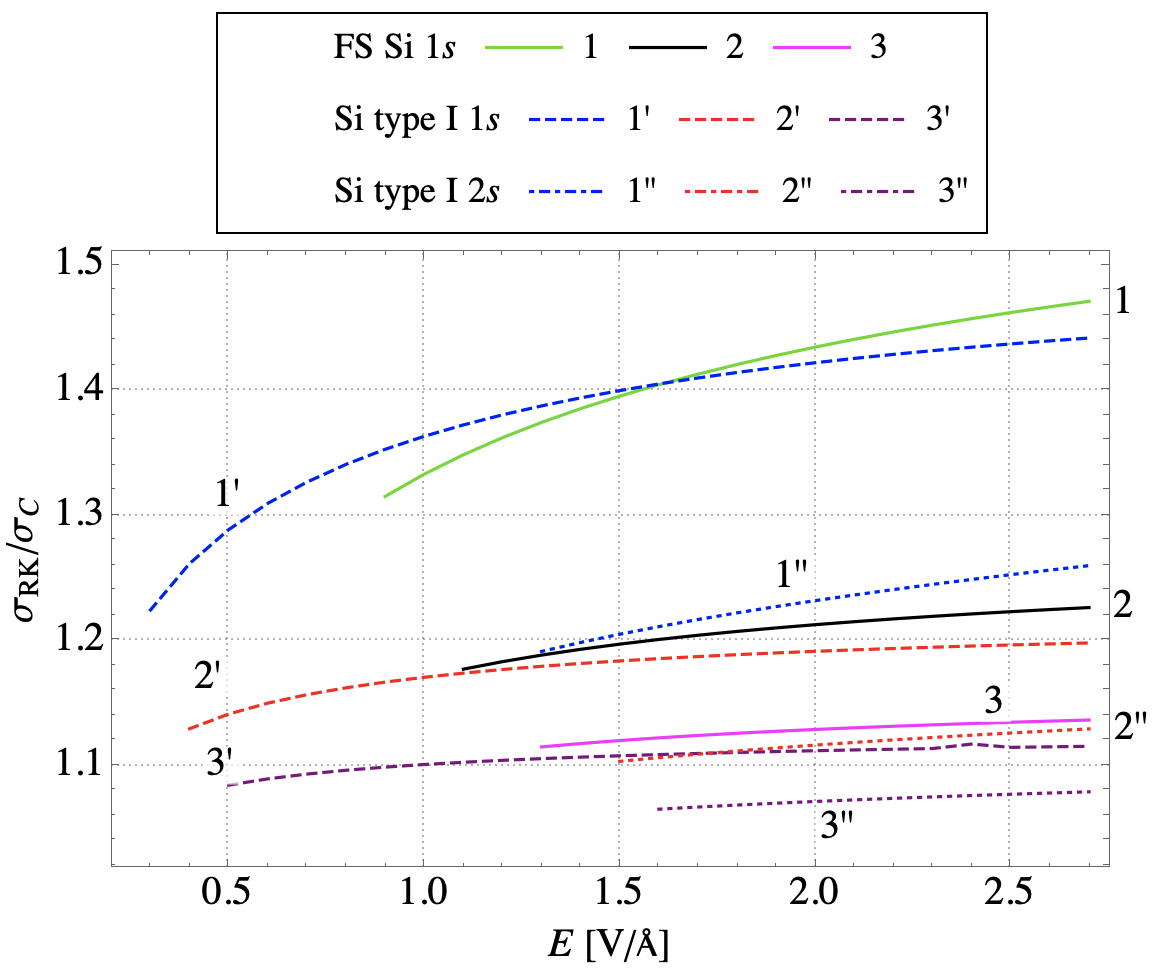}
\caption{Ratio $\sigma_{RK}$/$\sigma_C$ for indirect magnetoexcitons in FS Si and Si type I as a function of the electric field. For indirect magnetoexcitons in FS Si only ratio for 1$s$ state is shown since in state 2$s$ the linear dependence of $\Delta \mathcal{E}$ on $B^2$ is present only at very small values of the electric field. For indirect magnetoexcitons in Si type I ratio is shown for states 1$s$ and 2$s$. Data are plotted for cases where the number of hBN layers is $N$=1,2,3. For $N>3$ the ratio converges to one for all considered cases.} \label{si_I}
\end{figure}

We also analyze the dependence of the ratio $\sigma_{RK}$/$\sigma_C$, where $\sigma_{RK}$ and $\sigma_C$ are diamagnetic coefficients of indirect magnetoexcitons obtained using the RK and Coulomb potentials, respectively, on the external electric field and number of hBN layers. The corresponding results are presented in Fig. \ref{si_I} for the indirect magnetoexcitons in FS Si  for the state 1$s$ and Si type I for states 1$s$ and 2$s$. For the magnetoexcitons in FS Si in state 2$s$, there is no linear dependence of $\Delta \mathcal{E}$ on $B^2$, and the ratio is not presented. When the external electric field $E > 1.8 $ V/{\AA}, the DMCs for FS Si and Si type I are close and have the same qualitative dependence on $E$ and number of hBN layers. As a result, we can conclude that the diamagnetic coefficients can be tuned by the external electric field as well as by changing interlayer separation by changing the number of hBN layers.

\section{\label{sec:conclusion}Conclusion}

We study Rydberg states of direct and indirect magnetoexcitons in monolayers and double-layer heterostructures of Xenes in the presence of the external perpendicular to the layer electric and magnetic fields. We consider the freestanding silicene, germanene, and stanene monolayers and Si type I and II encapsulated by hBN monolayers and Xenes heterostructures. We have calculated the binding energies, the energy contribution from the electric and magnetic fields to the binding energy, and diamagnetic coefficients for magnetoexcitons in monolayer and heterostructure of Xenes. We solve the Schr\"{o}dinger equation with the magnetoexciton reduced mass dependent on the perpendicular electric field by numerical integration to obtain eigenvalues and eigenfunctions. For the direct magnetoexcitons, the Schr\"{o}dinger equation is solved using the Rytova-Kyldysh potential, and for the indirect excitons, the Schr\"{o}dinger equation is solved with both the RK and Coulomb potentials.

The contribution from the electric and magnetic fields to the binding energy of the magnetoexcitons in Rydberg states, 1$s$, 2$s$, 3$s$, and 4$s$, for freestanding and encapsulated monolayers as a function of the external perpendicular electric and magnetic fields is analyzed. The analysis of DMCs for magnetoexcitons in above mentioned systems and their dependence on the external electric field and number of hBN layers are presented as well. Only in monolayer magnetoexcitons had linear dependence of energy contributions on $B^2$ in states 1$s$, 2$s$, 3$s$, and 4$s$. The linear dependence of $\Delta \mathcal{E}$ on $B^2$ is not present for indirect magnetoexcitons in all states. The DMCs for the states 1$s$, 2$s$, 3$s$ and 4$s$ can be calculated only for direct magnetoexcitons in FS Si, FS Ge, FS Sn and Si type I. For indirect magnetoexcitons, DMCs can be calculated for states 1$s$ and 2$s$. As the number of layers that separate Xenes monolayers increases, the binding energies of Rydberg states decrease, and energy contribution from the electric and magnetic fields to the binding energy for given values of the electric field no longer has a linear dependence on $B^2$. A smaller effective mass $m$ (small external electric field) leads to a higher energy contribution from the magnetic field to the binding energy and potentially higher DMCs. However, small $m$ and high energy contribution from the magnetic field to the binding energy lead to absence of linear dependence of the energy contribution on $B^2$ and, therefore, the DMCs cannot be extracted.

The most bound direct magnetoexcitons are found in the FS Si monolayer. The magnetoexcitons in FS Si have the smallest effective and reduced masses out of all materials that we have examined. However, the material  constants of the encapsulating materials also play an important role in determining the properties of excitons in the monolayer and, therefore, the binding energy of the exciton as can be seen by examining the heterostructure. From our calculation, we can see that, by taking the magnetoexciton with parameters of FS monolayer and then the encapsulating monolayer with hBN, magnetoexcitons have smaller binding energies than magnetoexcitons in Si type I monolayer. The indirect magnetoexcitons are most bound in Si type I monolayer. So we conclude that, while the effective mass has the strong effect on the binding energy and the energy contribution from the magnetic field to the binding energy, the dielectric constants of the encapsulating materials also have a significant effect on the binding energy.

The analysis of Xenes parameters shows that Si type I has a smaller gap and Fermi velocity than Si type II, and, therefore, magnetoexcitons in Si type I monolayer have higher effective mass and binding energies than magnetoexcitons in Si type II. But the magnetoexcitons in Si type II has higher DMCs and higher energy contribution from the electric and magnetic fields to the binding energy in monolayer and heterostructure since the magnetoexciton in Si type II monolayer has smaller $m$ than the magnetoexciton in Si type I monolayer at the same value of the electric field. Thus, the importance of using physically accurate parameters of the material, since they affect the binding
energies and DMCs, is shown.

The comparison of the results for Xenes heterostructure obtained by using the Rytova-Keldysh and Coulomb potentials shows that $V_{RK}$ and $V_C$ converge as the number of layers increases. The binding energies, $\Delta\mathcal{E}$, and DMCs converge as the number of hBN layers increases. The energy contribution to the binding energy of the magnetoexciton calculated with $V_{RK}$ is higher than the energy contribution calculated using $V_{C}$ when only a few layers separate Xenes monolayers. The diamagnetic coefficients extracted from the binding energy of the magnetoexciton calculated using $V_{RK}$ are higher than when $V_C$ is used. However, in both cases, the DMCs converge as $E$ and the number of hBN layers increase. Therefore, the choice of the interaction potential affects the energy contribution from the electric and magnetic fields and DMCs (when the contribution is proportional to $B^2$), and it is important to use appropriate interaction potential for the system when only a few layers separate magnetoexciton containing monolayers.

Finally, we can conclude that external electric and magnetic fields can be effectively used for tuning their contribution to the binding energy of both direct and indirect magnetoexcitons in Xenes, while for the Xenes heterostructure the varying of the number of hBN separating layers provides the additional degree of freedom to tune the binding energy of indirect magnetoexcitons. The tunability of the  diamagnetic coefficients by the external electric field for both direct and indirect magnetoexcitons and by the change of the number of hBN separated Xenes layers for indirect magnetoexcitons is also demonstrated. Thus, the tunability of direct and indirect magnetoexcitons properties allows the design of electronic devices that can be manipulated by the external electric and magnetic fields and the number of hBN separating Xenes layers in heterostructure.

\textbf{Acknowledgments.}
This work is supported by the U.S. Department of
Defense under Gran No. W911NF1810433 and PSC-CUNY Award No. 62261-00 50.


\end{document}